\documentclass[11pt]{bmc_article}
\usepackage{amsmath}
\usepackage[utf8]{inputenc}
\usepackage{cite}
\usepackage{url}
\usepackage{ifthen}
\usepackage{multicol}
\usepackage{graphicx}

\newboolean{publ}

\newenvironment{bmcformat}{\fussy\setboolean{publ}{true}}{\fussy}

\newcommand{\ratet}{\mbox{$\alpha_t$}}
\newcommand{\rateinf}{\mbox{$\alpha$}}
\newcommand{\Npop}{\mbox{$N$}}

\newcommand{\mavgaa}{\mbox{$\langle m_{aa} \rangle$}}
\newcommand{\mavgnt}{\mbox{$\langle m_{nt} \rangle$}}

\newcommand{\mavg}{\mbox{$\langle m \rangle_{T}$}}

\newcommand{\mavgmono}{\mbox{$\langle m \rangle_{T,M}$}}
\newcommand{\mavgpoly}{\mbox{$\langle m \rangle_{T,P}$}}

\newcommand{\ddg}{\mbox{$\Delta\Delta G$}}

\newcommand{\dgfmin}{\mbox{$\Delta G_{f}^{\rm{min}}$}}
\newcommand{\dgf}{\mbox{$\Delta G_{f}$}}

\newcommand{\dgfextra}{\mbox{$\Delta G_{f}^{\rm{extra}}$}}

\newcommand{\avgFt}{\mbox{$\langle\mathcal{F}\rangle_t$}}
\newcommand{\avgF}{\mbox{$\langle\mathcal{F}\rangle$}}
\newcommand{\avgFmono}{\mbox{$\langle\mathcal{F}\rangle_M$}}
\newcommand{\avgFpoly}{\mbox{$\langle\mathcal{F}\rangle_{P}$}}
\newcommand{\ra}{\mbox{$\rightarrow$}}
\newcommand{\p}{\mbox{$\mathbf{p}$}}
\newcommand{\x}{\mbox{$\mathbf{x}$}}

\newcommand{\W}{\mbox{$\mathbf{W}$}}
\newcommand{\Wm}{\mbox{$\mathbf{W_m}$}}
\newcommand{\Wone}{\mbox{$\mathbf{W_1}$}}

\newcommand{\Wij}{\mbox{$W_{ij}$}}
\newcommand{\Wijm}{\mbox{$W_{ij,m}$}}
\newcommand{\Wjim}{\mbox{$W_{ji,m}$}}

\newcommand{\V}{\mbox{$\mathbf{V}$}}
\newcommand{\Vm}{\mbox{$\mathbf{V_m}$}}

\newcommand{\I}{\mbox{$\mathbf{I}$}}

\newcommand{\Viim}{\mbox{$V_{ii,m}$}}

\newcommand{\nuim}{\mbox{$\nu_{i,m}$}}

\newcommand{\psmall}{\mbox{$\mathbf{p_{o}}$}}
\newcommand{\pmono}{\mbox{$\mathbf{p_{M}}$}}
\newcommand{\xinf}{\mbox{$\mathbf{x_{\infty}}$}}
\newcommand{\xpoly}{\mbox{$\mathbf{x_{P}}$}}

\newcommand{\nusmall}{\mbox{$\langle \nu \rangle_{o}$}}

\newcommand{\nuinf}{\mbox{$\langle \nu \rangle_{\infty}$}}

\newcommand{\e}{\mbox{$\mathbf{e}$}}
\newcommand{\uhalf}{\mbox{[urea]$_{50}$}}
\newcommand{\thalf}{\mbox{$T_{50}$}}
\newcommand{\dgC}{\mbox{$^{\rm{o}}$C}}
\newcommand{\yzero}{\mbox{$\mathbf{y_0}$}}
\newcommand{\yone}{\mbox{$\mathbf{y_1}$}}
\newcommand{\ym}{\mbox{$\mathbf{y_m}$}}

\begin{document}
\begin{bmcformat}

\title{Evolution favors protein mutational robustness in sufficiently large populations} 

\author{Jesse D Bloom\correspondingauthor$^1$
\email{Jesse D Bloom\correspondingauthor - jesse.bloom@gmail.com} 
\and 
Zhongyi Lu$^1$
\email{Zhongyi Lu - lu07@caltech.edu} 
\and 
David Chen$^1$
\email{David Chen - davidc@caltech.edu} 
\and 
Alpan Raval$^2$
\email{Alpan Raval - alpan\_raval@kgi.edu} 
\and 
Ophelia S Venturelli$^1$
\email{Ophelia S Venturelli - opheliav@stanford.edu} 
and 
Frances H Arnold\correspondingauthor$^1$
\email{Frances H Arnold\correspondingauthor - frances@cheme.caltech.edu}
}

\address{
\iid (1) Division of Chemistry and Chemical Engineering, California Institute of Technology, Pasadena, California  91125, USA \\
\iid (2) Keck Graduate Institute of Applied Life Sciences and School of Mathematical Sciences, Claremont Graduate University, Claremont, CA  91711, USA
}

\maketitle

\begin{abstract}
\paragraph*{Background:} An important question is whether evolution favors properties such as mutational robustness or evolvability that do not directly benefit any individual, but can influence the course of future evolution.  Functionally similar proteins can differ substantially in their robustness to mutations and capacity to evolve new functions, but it has remained unclear whether any of these differences might be due to evolutionary selection for these properties.
\paragraph*{Results:}  Here we use laboratory experiments to demonstrate that evolution favors protein mutational robustness if the evolving population is sufficiently large.  We neutrally evolve cytochrome P450 proteins under identical selection pressures and mutation rates in populations of different sizes, and show that proteins from the larger and thus more polymorphic population tend towards higher mutational robustness.  Proteins from the larger population also evolve greater stability, a biophysical property that is known to enhance both mutational robustness and evolvability.  The excess mutational robustness and stability is well described by existing mathematical theories, and can be quantitatively related to the way that the proteins occupy their neutral network.  
\paragraph*{Conclusions:} Our work is the first experimental demonstration of the general tendency of evolution to favor mutational robustness and protein stability in highly polymorphic populations.  We suggest that this phenomenon may contribute to the mutational robustness and evolvability of viruses and bacteria that exist in large populations.
\end{abstract}

\ifthenelse{\boolean{publ}}{\begin{multicols}{2}}{}

\section*{Background}
Proteins are quite tolerant of mutations, allowing evolution to produce highly diverged sequences that fold to similar structures and perform conserved biochemical functions~\cite{Zuckerkandl1965,Lesk1980}.  However, proteins with nearly identical structures and functions may differ in their robustness to mutation~\cite{Bloom2005,Besenmatter2007,Bloom2006}, as well as in their capacity to acquire new functions~\cite{Bloom2006}.  The fact that mutational robustness and evolvability can vary among the functionally equivalent proteins produced by natural sequence divergence makes these properties important hidden dimensions in evolution --- direct selection for protein function is blind to them, yet they can play a crucial role in enabling future evolution.  Whether the evolutionary process somehow promotes the acquisition of mutational robustness and evolvability therefore remains a major question~\cite{Lenski2006,Sniegowski2006,Wagner2005book}.\pb

Previous experiments have identified several specific evolutionary conditions that can affect mutational robustness.  For example, genetic complementation decreases the mutational robustness of viruses~\cite{Montville2005}, while high mutation rates favor mutational robustness in simulated digital organisms~\cite{Wilke2001b}.  However, theory~\cite{vanNimwegen1999} makes the much broader --- and heretofore experimentally untested --- prediction that extra mutational robustness will arise quite generally in sufficiently large populations.  This prediction cannot be understood in the standard framework of Kimura's neutral theory~\cite{Kimura1983}, since one of the usual assumptions of the neutral theory is that mutational robustness is constant. (Although Takahata~\cite{Takahata1987} treated the consequences of stochastically fluctuating neutrality on the molecular clock, he did not describe how mutational robustness might change systematically during evolution.)  However, changes in mutational robustness can be described by envisioning evolution as occurring on neutral networks, or sets of functionally equivalent proteins that are connected by single mutational steps~\cite{Smith1970,Lipman1991,Huynen1996,Bornberg-Bauer1999}.  In a seminal theoretical analysis of evolution on neutral networks,  van Nimwegen and coworkers~\cite{vanNimwegen1999} predicted that the extent of mutational robustness should depend on the degree of population polymorphism.  Here we briefly summarize their reasoning, since it motivates our experimental work.  We also refer the reader to chapter 16 of \cite{Wagner2005book}, which contains an excellent explanation of the densely mathematical work of van Nimwegen and coworkers~\cite{vanNimwegen1999}.\pb

If an evolving population is mostly monomorphic, then each mutation is either lost or goes to fixation before another mutation occurs.  The population is therefore usually clustered at a single genotype and rarely experiences mutations, meaning that selection does not distinguish between genotypes of different mutational robustness.  All nodes of the neutral network are thus equivalent and will be occupied by the population with equal probability~\cite{vanNimwegen1999}.  On the other hand, a highly polymorphic population is always spread across many nodes of the neutral network.  When mutations occur, the members of the population at highly connected nodes have a better chance of surviving, causing them to be favored by evolution and increasing the average mutational robustness ~\cite{vanNimwegen1999,Bornberg-Bauer1999,Wilke2001,Taverna2002b,Xia2004b}.  Specifically, a highly polymorphic population occupies each node with a probability proportional to its eigenvector centrality~\cite{vanNimwegen1999,Bornberg-Bauer1999}, a measure of how connected it is to other connected nodes (a variant of eigenvector centrality is used by Google's PageRank algorithm to rank a webpage's importance in the network of internet links~\cite{Brin1998}).  Figure \ref{fig:model}A illustrates how mostly monomorphic and highly polymorphic populations are predicted to occupy a neutral network.  For proteins, changes in neutral network occupancy should be manifested by changes in thermodynamic stability~\cite{Bloom2007}, with proteins from highly polymorphic populations predicted to be more stable than their counterparts from mostly monomorphic populations (Figure \ref{fig:model}B).  Note that the extent of polymorphism depends on the product of the mutation rate and population size, meaning that protein populations of different sizes are predicted to evolve to different levels of mutational robustness and stability even if they experience the same mutation rate.\pb

\section*{Results and Discussion}
\subsection*{Design of neutral evolution experiment}
To test whether high population polymorphism drives an increase in mutational robustness and protein stability, we performed laboratory evolution experiments on cytochrome P450 proteins.  The basic idea was to neutrally evolve P450s under a constant selection pressure in populations that were either monomorphic or highly polymorphic, and observe whether the proteins evolved to different levels of mutational robustness and stability.  The evolution experiments started with a P450 BM3 heme domain that had been engineered to hydroxylate 12-$p$-nitrophenoxydodecanoic acid (12-pNCA) \cite{Cirino2003}.  We imposed the selection criterion that \textit{Escherichia coli} cells expressing the P450 had to yield lysate with enough active enzyme to hydroxylate a specified amount of 12-pNCA in 40 minutes.  This criterion roughly corresponds to the case in which an enzyme must catalyze a biochemically relevant reaction at some minimal level in order for its host to survive.  Note that other properties such as stability and expression level can vary freely, provided that the criterion for total activity is met.\pb

The properties of a neutrally evolving protein eventually ``equilibrate,'' much as the properties of an isolated physical system under some macroscopic constraint tend towards the values that maximize the system's internal entropy.  For proteins, this usually means that stability, expression, and activity drift towards their lowest tolerable values, since the vast majority of random sequences do not encode stable, well-expressed enzymes (that is, natural selection must work against sequence entropy to maintain a functional protein)~\cite{Bloom2007,Taverna2002}.  The initial P450 had been engineered for maximal activity~\cite{Cirino2003}, meaning that it was not equilibrated to the more mild selection criterion of the experiments.  We therefore neutrally evolved this initial P450 for 16 generations, introducing random mutations with error-prone PCR and retaining all mutants that met the selection criterion for total activity on 12-pNCA.  The procedure used for this equilibration evolution was similar to that for the polymorphic neutral evolution described below.  As expected, expression, stability, and activity all dropped during the equilibration evolution.   At the end of the equilibration evolution, we chose a single sequence as the parent for the neutral evolution experiments.  The gene encoding this parent sequence contained 29 nucleotide mutations and 13 amino acid mutations relative to the initial P450 (Additional File \ref{add:parent_P450_R1-11_sequence}).\pb

We used this parent gene to begin three parallel sets of neutral evolution experiments, which we named ``monomorphic,'' ``polymorphic,'' and ``unselected'' (Figure \ref{fig:experiment}).  The monomorphic experiments capture the case where the population moves as a single entity, the polymorphic experiment captures the case where the population spreads across many sequences, and the unselected experiments show how the gene evolves in the absence of selection for protein function.  In all experiments, at each generation we used error-prone PCR to introduce an average of 1.4 nucleotide mutations per P450 gene (Table \ref{tab:mutspectrum}).  The mutant genes were ligated into a plasmid and transformed into \textit{E. coli}~\cite{Barnes1991}, and transformants were selected using the plasmid's antibiotic resistance marker.  For the unselected case, we randomly picked one of the mutants, recovered the mutant gene with a plasmid mini-prep, and used this mutant as the template for the next generation of error-prone PCR.  We performed four independent replicates of unselected evolution, evolving each for 12 generations.\pb

For the monomorphic and polymorphic populations, we imposed the selection criterion that the P450s hydroxylate 12-pNCA with at least 75\% of the total activity of the original parent gene.  We expressed the P450s in \textit{E. coli}, and then assayed the cell lysates for activity in a high-throughput 96-well plate format.  The total amount of product produced by 80 $\mu$l of clarified lysate in 40 minutes was compared to the median of four control wells containing the original parent P450 to determine if the mutant met the selection criterion.  The only difference between the monomorphic and polymorphic experiments was the size of the evolving populations.  In the monomorphic limit, each mutation is either lost or goes to fixation before the next occurs.  We enforced this evolutionary dynamic by holding the population size to a single protein sequence, similar to the ``blind ant'' random walk of \cite{vanNimwegen1999}.  At each generation, we assayed a single mutant.  If this mutant met the selection criterion, then it was carried over to the next generation, corresponding to a neutral mutation going to fixation.  If the mutant failed the selection criterion, then the population stayed at the previous sequence for the next generation, corresponding to a mutation lost to selection.  If all of the mutants assayed had zero or one mutations, then this protocol would correspond exactly to the equations of \cite{vanNimwegen1999,Bloom2007}.  However, in order to achieve appreciable sequence evolution on a laboratory time scale, we used a mutation rate that sometimes produced multiple mutations in a generation.  We mathematically describe this situation in the Mathematical Appendix; here we simply note that it is possible to think of each generation as introducing a single mutational event rather than a single mutation.  We performed 22 independent replicates of monomorphic evolution, evolving each for 25 generations.\pb

In the polymorphic limit, the population spreads across many sequences.  To implement this experimentally, we assayed 435 mutants at each generation.  The selection criterion was used to classify each mutant as functional or nonfunctional.  In neutral evolution, all functional mutants reproduce with equal probability.  We therefore pooled equal volumes of stationary-phase cultures of each functional mutant and recovered the pooled genes with a mini-prep.  The polymorphic evolution experiment therefore approaches the equations of \cite{vanNimwegen1999,Bloom2007}, again with the exception that a sequence may undergo multiple mutations at a single generation.  We give the equations describing this situation in the Mathematical Appendix.  Since the population evolves deterministically in the polymorphic limit~\cite{vanNimwegen1999,Bloom2007}, a single replicate was performed.  Because mutations accumulate more rapidly in the polymorphic experiments than the monomorphic ones, we evolved the polymorphic population for 15 generations rather than 25. \pb

\subsection*{Mutations and mutational robustness}
Figure \ref{fig:mutations} shows how mutations accumulated during the course of the neutral evolution experiments (full data are in Table \ref{tab:evolutionsummary} and Additional File \ref{add:evolution_data}).  Since the unselected protein populations evolve without constraint, mutations accumulate at the same rate at which they are introduced by error-prone PCR, 1.4 nucleotide mutations per generation.  Because selection eliminates mutations that disrupt P450 activity, mutations accumulate more slowly in the monomorphic and polymorphic populations.  Mutations accumulate more rapidly in the polymorphic population than in the monomorphic populations.  This difference in rates is predicted by the equations in the Mathematical Appendix to be a consequence of the fact that the polymorphic population is more mutationally robust, and so can tolerate more of the possible mutations.  \pb

To test directly whether the polymorphic population evolves higher average mutational robustness, we measured the fraction of 435 random mutants that met the selection criterion.  Figure \ref{fig:robustness} shows that the polymorphic population neutrally evolved to a markedly higher mutational robustness than the monomorphic populations, with $50 \pm 2$\% of the final polymorphic population mutants continuing to function versus $39 \pm 2$\% for the final monomorphic populations (Chi-square $P$-value of $10^{-3}$ that these values are significantly different). The only difference between the two types of populations was their size, so evolution has clearly favored mutational robustness in the larger and thus more polymorphic population.  This finding represents the first experimental support for the prediction that highly polymorphic populations evolve excess mutational robustness~\cite{vanNimwegen1999}.\pb

Theory predicts that the excess mutational robustness of a highly polymorphic protein population comes from increased protein stability~\cite{Bloom2007}.  Because the P450 variants unfold irreversibly, an equilibrium thermodynamic stability \dgf\ cannot be measured.  We therefore determined stability to irreversible thermal and chemical denaturation, two highly correlated measures of P450 stability that have previously been shown to contribute to mutational robustness~\cite{Bloom2006} (see Additional Files \ref{add:thermostabilities}, \ref{add:urea_stabilities}, and \ref{add:T50_U50_correlation}).  Figure \ref{fig:stabilities_expression} shows that proteins from the polymorphic population were in fact more stable than their counterparts from the monomorphic population.   We also observed that proteins in the polymorphic population tended to accumulate to higher levels in \textit{E. coli} (Figure \ref{fig:stabilities_expression}).  Elevated expression could be a byproduct of increased stability, or it could independently increase mutational robustness by allowing the proteins to better tolerate mutations that decrease codon adaptation or reduce folding efficiency.  It is possible that additional unrecognized biophysical factors also contributed to the excess mutational robustness of the polymorphic population, but no such factors were immediately obvious.\pb

\subsection*{Interpretation in terms of the P450 neutral network}
The higher mutational robustness of the polymorphic population is due to the fact that it occupies the P450 gene neutral network differently than the monomorphic populations.  Measurements from the evolution experiments can therefore be used to infer basic properties of the underlying neutral network of P450 genes, as originally noted by van Nimwegen and coworkers~\cite{vanNimwegen1999}.  In the Mathematical Appendix, we derive approximations for the normalized principal eigenvalue \nuinf\ and the normalized average connectivity \nusmall\ of the neutral network, where in both cases the normalization is obtained by dividing by the network coordination number.  We obtain $\nuinf = 0.51$ and $\nusmall = 0.35$ for the P450 gene neutral network.  Our ability to consistently estimate these two parameters from four different experimental measurements supports the idea that the theory that we elaborate in the Mathematical Appendix appropriately describes the experiments.  The difference between \nuinf\ and \nusmall\ is a measure of the extent to which some P450 neutral network nodes have more connections than others.  We note that \nuinf\ is approximately equal to the exponential decline parameter for the asymptotic decline in the fraction of functional mutants with increasing numbers of random nucleotide mutations~\cite{Bloom2005,Shafikhani1997,Guo2004} (see Mathematical Appendix).  Previous studies looking at this exponential decline have reported $\nuinf = 0.7$ for subtilisin~\cite{Shafikhani1997}, $\nuinf = 0.7$ for 3-methyladenine DNA glycosylase~\cite{Guo2004}, and $\nuinf = 0.7$ - $0.8$ for TEM1 $\beta$-lactamase~\cite{Bloom2005}.  These comparisons suggest that P450 has a sparser neutral network (smaller \nuinf) than these other proteins.  We suspect, however, that these earlier studies (one of which is our own) overestimate \nuinf\ due to insufficient equilibration of the starting sequence.  We believe that the approach of the current work is more accurate for determining \nuinf\ because the measurements are made after many mutations have equilibrated the initial sequence.  This approach could be used in future experiments to compare the neutral network connectivities of proteins from different families.\pb

\section*{Conclusions}
We have demonstrated that neutral evolution favors more mutationally robust proteins when the evolving population is highly polymorphic.  Strikingly, the excess mutational robustness is due only to population polymorphism, and so will arise in any population of sufficiently large size.  Our work is the first experimental demonstration of this phenomenon, which is predicted to occur quite generally in neutrally evolving proteins and nucleic acids~\cite{vanNimwegen1999}.  Furthermore, we were able to identify one of the biophysical factors underlying the increase in mutational robustness by showing that proteins from the highly polymorphic population are more stable.  We recognize that evolution in a biological context will be more complex.  In our experiments, fitness was the P450's ability to be expressed in active form by bacteria grown to saturation in an environment with plentiful nutrients.  Biological fitness, however, depends on numerous additional and subtle effects such as the metabolic costs of synthesis or the burdens imposed by misfolded molecules.  Some mutations that are neutral in the experiments may therefore have deleterious effects in a biological setting~\cite{DePristo2005}.  The experiments nonetheless capture the overriding constraint that proteins retain their biochemical functions.  Our success in quantitatively explaining the results supports the notion that important aspects of protein evolution can be described simply in terms of mutational effects on stability~\cite{Bloom2007,DePristo2005}.  \pb

An obvious question is whether evolution in nature favors mutational robustness by the process we have demonstrated.  Whether natural populations will neutrally evolve mutational robustness depends on whether they are sufficiently polymorphic, which will be the case if the product of their effective population size \Npop\ and per protein per generation mutation rate $\mu$ is much greater than one~\cite{vanNimwegen1999,Kimura1983}.  Accurately estimating $\Npop\mu$, which is closely related to the widely used parameter $\theta$ in population genetics, for natural populations is difficult~\cite{Plotkin2006,Berg1996} (note that since mutational robustness is a protein-wide property, the relevant mutation rate is per protein, which is $\approx 10^2$ to $10^3$ larger than the per codon mutation rate).  For humans and other multicellular organisms,  $\Npop\mu$ is probably too small~\cite{Lynch2003} for their proteins to neutrally evolve mutational robustness.   But estimates~\cite{Lynch2003,Hartl1994} place $\Npop\mu \approx 10$ to $100$ for typical-length proteins in bacteria, and it is probably much higher for many viruses~\cite{Moya2004,Vignuzzi2006}.  It is therefore likely that many viral and some bacterial proteins have neutrally evolved extra mutational robustness.\pb

The neutral evolution of protein mutational robustness may also contribute to adaptive evolution.  Experiments have shown that extra stability increases a protein's evolvability by allowing it to tolerate a wider range of functionally beneficial but destabilizing mutations~\cite{Bloom2006}.  A similar phenomenon seems to occur in natural evolution, where functionally neutral but stabilizing mutations can play a key role in adaptive evolution by counterbalancing the destabilizing effects of other functionally beneficial mutations~\cite{Wang2002}.  Viruses and perhaps bacteria may thus benefit from large population sizes and high mutation rates that drive an increase in the mutational robustness and stability of their proteins, which in turn enhances the capacity of these proteins to rapidly change their sequences and evolve new functions.  \pb 

\section*{Methods}
\subsection*{Equilibration evolution of the P450 protein}
We began with a 21B3 P450 peroxygenase that had been engineered for highly efficient hydroxylation of 12-pNCA~\cite{Cirino2003} (sequence shown in Additional File \ref{add:initial_P450_21B3_sequence}).  This P450 was not well equilibrated to the constant selection criterion that we planned to impose, since it had substantially higher total activity.  We therefore neutrally evolved it for 16 generations in order to create P450s that were better equilibrated to the selection criterion.  We evolved two parallel populations, which we named R1 and R2.  The procedure was exactly identical to that described below for the polymorphic evolution with the following exceptions:
\begin{itemize}
\item Starting sequence: the starting sequence for the equilibration evolution was the 21B3 sequence.
\item Population size: each of the two equilibration evolution populations had a size of 174 sequences rather than the 435 used for the polymorphic evolution.
\item Selection criterion: the sequences were required to have at least 75\% of the total activity of the 21B3 P450.
\item Mutation rate: the mutation rate for the equilibration evolution was much higher than for the polymorphic evolution.  The error-prone PCR protocol used 200 $\mu$M manganese chloride (MnCl$_2$), rather than the 25 $\mu$M used for the polymorphic evolution.  We estimate that this error-prone PCR protocol introduced $\approx 4$ nucleotide mutations per P450 gene at each generation during the equilibration evolution.
\end{itemize}
We performed 16 generations of equilibration evolution, and then randomly selected 23 functional mutants from each of the R1 and R2 populations (Additional File \ref{add:equilibration_mutations}).  We picked one of these mutants, R1-11, for use as the parent for the neutral evolution experiments. \pb

\subsection*{Detailed protocol for evolution experiments}
We began with the R1-11 P450 BM3 heme domain variant (sequence in Additional File \ref{add:parent_P450_R1-11_sequence}) cloned into the pCWori~\cite{Barnes1991} plasmid with a 5' \textit{BamH1} and 3' \textit{EcoR1} site as described in \cite{Bloom2006}.  The cloning primers were \textit{pCWori\_for} ({\footnotesize 5'-GAAACAGGATCCATCGATGCTTAGGAGGTCAT-3'} and \textit{pCWori\_rev\_clone} ({\footnotesize 5'-GCTCATGTTTGACAGCTTATCATCG-3'}).  We used error-prone PCR  to generate mutants, taking great care to make the error-prone PCR protocol repeatable by using a relatively small number of thermal cycles.  This was both to control the mutation rate by ensuring that the reaction did not saturate the reagents (which would cause the number of doublings to become sensitive to the initial template concentration), and to avoid the PCR-based recombination events which can occur during with the last few thermal cycles of PCR reactions~\cite{Meyerhans1990,Kanagawa2003}.  The PCR reactions were 100 $\mu$l in volume, and contained $\approx$ 13 ng of plasmid template (corresponding to $\approx 3$ ng of template gene), 7 mM magnesium chloride MgCl$_2$, 1 $\times$ Applied BioSystems PCR Buffer II without MgCl$_2$, 25 $\mu$M MnCl$_2$, 0.5 $\mu$M \textit{pCWori\_for} primer, 0.5 $\mu$M \textit{pCWori\_rev} primer, 200 $\mu$M of dATP and dGTP, 500 $\mu$m of dTTP and dCTP, and 5 units of Applied Biosystems AmpliTaq polymerase.  The reactions were run on the BLOCK setting of a MJ Research PCR machine with a program of 95\dgC\ for 2 minutes, then 15 cycles of (95\dgC\ for 30 seconds, 57\dgC\ for 30 seconds, 72\dgC\ for 90 seconds), and then cooling to 4\dgC.  This protocol yielded roughly 1-1.5 $\mu$g of product gene (as quantified by gel electrophoresis versus a known standard), for a PCR efficiency of $\approx 0.5$.  Sequencing the unselected populations at the end of the experiment indicated that this protocol introduced an average of $1.4 \pm 0.2$ nucleotide mutations, with the nucleotide error-spectrum shown in Table \ref{tab:mutspectrum}.  Because the number of PCR doublings is large compared the average mutation rate, the distribution of mutations among sequences should be well-described by the Poisson distribution~\cite{Sun1995,Drummond2005c}.\pb

The mutant genes from the error-prone PCR were purified over a ZymoResearch DNA clean and concentrator column, and digested at 37\dgC\ with \textit{EcoR1} and \textit{BamH1}.  The digested genes were then purified from an agarose gel with ZymoResearch DNA gel extraction columns, and ligated into pCWori plasmid that had been digested with \textit{BamH1} and \textit{EcoR1} and dephosphorylated.  The ligations were transformed into electro-competent catalase-free \textit{Escherichia coli}~\cite{Barnes1991} (the catalase is removed because it breaks down the hydrogen peroxide utilized by the P450 peroxygenase), plated on Luria Broth (LB) plates containing 100 $\mu$g/ml of ampicillin to select for the plasmid's antibiotic resistance marker, and grown at 37\dgC.  Transformation of a control ligation reaction without any digested gene yielded at least 100-fold fewer colonies, indicating that the rate of plasmid self-ligation was less than one percent.\pb

Individual mutant colonies from the plates were picked into 96-well 2 ml deep-well plates containing 400 $\mu$l of LB supplemented with 100 $\mu$g/ml ampicillin.  Each plate contained four parental control wells with cells carrying the parent R1-11 gene, four null control wells with cells carrying the pCWori plasmid without a P450 gene, and a non-inoculated well to check for contamination.  For the polymorphic population, we picked five such plates with all 87 other wells containing different mutants for a total population size of $5 \times 87 = 435$ mutants.  For the 22 monomorphic populations (we began with 24 populations but two had to be discarded due to contamination), we picked a single colony for growth and screening.  For the unselected populations we picked a single colony for growth without screening.  The LB deep-well plates were grown for 16-20 hours at 30\dgC, 210 revolutions per minute (rpm), and 80\% relative humidity in a Kuhner humidified shaker.  To express the P450 mutants, we prepared 2 ml deep well plates containing 400 $\mu$l per well of terrific broth (TB) supplemented with 200 $\mu$M isopropyl $\beta$-D-thiogalactoside (IPTG), 100 $\mu$g/ml ampicillin, and 500 $\mu$M of $\delta$-aminolevulinic acid.  We used a pipetting robot inoculated these TB plates with 100 $\mu$l from the LB plates.  We stored the LB deep-well plates at 4\dgC, and grew the TB deep-well plates in the humidified shaker at 30\dgC, 210 rpm, and 80\% relative humidity for 22-24 hours.  After this growth, the cells were harvested by centrifuging the TB deep-well plates at $4000 \times$g for 5 minutes and discarding the liquid.  The cell pellets were flash-frozen in liquid nitrogen to aid in cell lysis.\pb

To lyse the cells for the assays, we resuspended the cell pellets in 300 $\mu$l of  100 mM [4-(2-hydroxyethyl)-1-piperazinepropanesulfonic acid] (EPPS) (pH 8.2) with 0.5 mg/ml lysozyme and 4 units/ml of deoxyribonuclease by pipetting 40 times with the pipetting robot.  We then incubated the plates at 37\dgC\ for 30 minutes, again resuspended with the pipetting robot, and put back at 37\dgC\ for an addition 30 minutes.  We then pelleted the cell debris by centrifugation at 6000$\times$g for 5 minutes at 4\dgC.  The pipetting robot was used to dispense 80 $\mu$l of the clarified lysate into 96-well microtiter plates (Rainin).  We prepared a 6$\times$ stock of 1.5 mM 12-pNCA in 36\% dimethyl sulfoxide (DMSO) and the EPPS buffer (the 12-pNCA was stored in the DMSO solution and combined with the buffer immediately before use).  We used a multichannel pipette to add 20 $\mu$l of this substrate stock to each well of the microtiter plate.  We briefly mixed the plates with ``shake'' setting of a 96-well plate spectrophotometer, and read an absorbance baseline at 398 nm.  We then immediately added 20 $\mu$l of a freshly prepared solution of 24 mM hydrogen peroxide in the EPPS buffer to initiate the reaction, and mixed again.  The final reaction conditions were therefore the EPPS buffer with 6\% DMSO, 4 mM hydrogen peroxide, and 250 $\mu$M 12-pNCA.  After 40 minutes we quantified the amount of enzymatic product by the increase in absorbance at 398 nm.  This absorbance increase is due to the 4-nitrophenolate molecule released after the P450 hydroxylates the twelfth carbon of the 12-pNCA molecule~\cite{Cirino2003}.  To score the mutants as functional or nonfunctional, we compared their gain in absorbance minus the median null control reading to that of the median parental control reading minus the median null control reading.  All mutants that had at least 75\% of the parental gain were scored as functional, all other mutants were scored as nonfunctional.\pb

We used the information from these assays to select the parents for the next generation.  For the unselected population we did not require the mutants to be functional, so the selected mutant was used to start a 4 ml culture of LB with 100 $\mu$g/ml ampicillin, and the plasmid DNA was harvested with a mini-prep.  This plasmid DNA was used as the template for the next round of error-prone PCR.  Therefore, after the first generation the four unselected replicates diverged into four separate error-prone PCR reactions.  These unselected replicates were evolved for a total of twelve generations, and were sequenced at every third generation.  \pb

For the polymorphic population, all mutants that were functional contributed an equal amount of plasmid DNA as template for the next generation.  In order to do this, we collected 50 $\mu$l of the culture from the LB deep-well plate for each mutant that was scored as functional.  All of these LB aliquots were pooled, and then the plasmid DNA was collected with a mini-prep.  The pool of plasmid DNA was used as template for the next generation's error-prone PCR reactions.  We performed 15 generations of evolution for this polymorphic population.  Note that at each generation we are assaying 435 mutants as part of the evolutionary procedure, so this provides information on mutational robustness.  At every third generation, we also selected a random sample of functional mutants for sequencing.  After 15 generations, we randomly selected 22 mutants for stability measurements and sequencing analysis.  The random selections were made from all functional mutants with the Python computer language random number generator.\pb

For the monomorphic populations, at each generation we assayed just a single mutant.  If that mutant was nonfunctional, then at that generation the population stayed at its original sequence.  In that case, for the next generation we simply picked a new mutant from the previous generation's plate of transformed mutants.  If the mutant we screened was functional, then that mutant represented the new population.  We therefore grew a 4 ml LB culture with 100 $\mu$g/ml of ampicillin, and collected the plasmid DNA with a miniprep.  That plasmid DNA was then used as the template for the next generation's error-prone PCR reaction.  We thus had 22 (actually 24 before 2 were contaminated) independent monomorphic populations that were being evolved in parallel.  Each was evolved for 25 generations, and at the end of these 25 generations we measured the stability of the final sequence of each population.  Each time an assayed mutant was functional, we sequenced the new P450 gene.  We also measured the average mutational robustness of the monomorphic populations at every fifth generation.  To do this, we did a pooled mini-prep of equal volumes of LB cultures of all 22 replicates to obtain a equal mix of plasmid DNA.  We then performed error-prone PCR on this mix, and assayed 435 mutants to measure the fraction functional.  Full neutral evolution data are in Additional File \ref{add:evolution_data}.\pb

\subsection*{Test for recombination during error-prone PCR}
During the polymorphic population evolution, we performed error-prone PCR on a mix of different plasmids.  It is common for PCR on mixed templates to lead to recombination events during the reaction~\cite{Meyerhans1990,Kanagawa2003}.  We attempted to reduce this recombination by using a small number of thermal cycles.  However, in order to test for recombination, we analyzed the sequences of the final 22 selected members of the polymorphic population.  There are a variety of statistical tests to detect recombination in a set of sequences.  A comparison of these tests by Posada~\cite{Posada2002} found that the Max-Chi$^2$ method developed by John Maynard Smith~\cite{Smith1992} performs well.  A publicly available implementation of this method~\cite{Piganeau2004} is at \url{http://www.lifesci.sussex.ac.uk/CSE/test/maxchi.php}.  We used this implementation to analyze the 22 final polymorphic sequences, and the resulting $P$-value was 0.29 after 100 random permutations, indicating that there is not significant  recombination.  \pb

\subsection*{Measurement of P450 stabilities}
We measured the stabilities to both irreversible thermal and irreversible urea denaturation of the final (generation 25) member of each monomorphic population, as well as of the 22 randomly selected members of the polymorphic population.   As discussed in the Supplementary Information of \cite{Bloom2006}, cytochrome P450 BM3 heme domains (and indeed most P450s) denature irreversibly, forcing us to use resistance to irreversible denaturation to quantify protein stability.  The first stability measure is the \thalf, defined as the temperature at which half of the protein is denatured after a 10 minute incubation.  The second stability measure is the \uhalf, defined as the urea concentration at which half of the protein denatures after a 4 hour room-temperature incubation.  Each set of measurements (those of \thalf\ and \uhalf) was performed on all of the mutants in the same day, and each mutant was treated identically.  Therefore, it is possible to make accurate comparisons of the relative values of the measurements within the data set.  However, the absolute values of the \thalf\ and \uhalf\ values may be less accurate.  Therefore, care should be taken in comparing  the absolute value of these measurements to those of other studies (such as \cite{Bloom2006}).\pb

Both the \thalf\ and \uhalf\ measurements were performed in clarified cell lysate.  The protein was expressed using catalase-free \textit{E. coli}~\cite{Barnes1991} containing the encoding gene on the IPTG inducible pCWori~\cite{Barnes1991} plasmid.  We used freshly streaked cells to inoculate 2 ml cultures of LB supplemented with 100 $\mu$g/ml of ampicillin, and grew these starter cultures overnight with shaking at 37\dgC.  We then used 0.5 ml from these starter cultures to inoculate 1 L flasks containing 200 ml of TB supplemented with 100 $\mu$g/ml of ampicillin.  The TB cultures were grown at 30\dgC\ and 210 rpm until they reached an optical density at 600 nm of $\approx$0.9, at which point IPTG and $\delta$-aminolevulinic acid were added to a final concentration of 0.5 mM each.  The cultures were grown for an additional 19 hours, then the cells were harvested by pelletting 50 ml aliquots at 5,500 g and 4\dgC\ for 10 min, and stored at -20\dgC.  To obtain clarified lysate, each pellet was resuspended in 8 ml of 100 mM EPPS (pH 8.2) and lysed by sonication, while being kept on ice.  The cell debris was pelleted by centrifugation at 8,000 g and 4\dgC\ for 10 minutes, and the clarified lysate was decanted and kept on ice.  \pb

For the \thalf\ measurements, 125 $\mu$l of clarified lysate from a single mutant was added to all 12 wells in a row of a 96-well hard-shell thin-wall microplate (MJ Research).  The plate was heated for 10 minutes using the gradient method of an Eppendorf Mastercycler gradient PCR machine, with the gradient set at either 33\dgC-45\dgC\ or 46\dgC-58\dgC\ (each mutant was exposed to both of these gradients), the machine on the BLOCK setting, and the heated lid set to 75\dgC\ with the lid WAIT option.  The plate was then cooled to 4\dgC, removed from the PCR machine, and centrifuged at 5,500 g and 4\dgC\ for 5 minutes to pellet any debris.  A pipetting robot was used to dispense 80 $\mu$l of the supernatent into a 96-well microtiter plate (Rainin), and the amount of remaining properly folded P450 was quantified from the carbon monoxide difference spectrum as described below.  The \thalf\ values were determined by fitting sigmoidal curves the percent of remaining protein as shown in Additional File \ref{add:thermostabilities}.  Our ability to accurately compare \thalf\ values within the data set requires that each well in a given column of the gradient PCR machine be at the same temperature.  We used a thermocouple to measure the temperature of the wells with the machine lid open, and confirmed that the wells were within a few tenths of a degree of the same temperature.  Further evidence for the consistency of our \thalf\ values comes from the fact that two independent measurements of the \thalf\ for our R1-11 parent yielded values that differed by only 0.1\dgC.  However, the absolute values of the measured temperatures are less accurate.  Thermocouple measurements indicated that, with the machine lid open, the wells were $\approx$ 1\dgC\ cooler than the indicated temperature.  We were unable to ascertain the temperatures with the heated lid closed, but based on comparisons water bath measurements, the temperatures with the lid closed slightly exceeded the indicated temperatures.  \pb

For the \uhalf\ measurements, 125 $\mu$l of the clarified lysate from a single mutant was added to all 12 wells in a row of a 96-well microtiter plate.  A pipetting robot was then used to add and mix 125 $\mu$l of a 2X solution of urea in 100 mM EPPS (pH 8.2) so that each subsequent column had a higher concentration of urea, and so that the final urea concentrations were those shown in Additional File \ref{add:urea_stabilities}.  The plates were left on the bench at room temperature for 4 hours, and the amount of remaining properly folded P450 was quantified from the carbon monoxide difference spectrum as described below.  The \uhalf\ values were determined by fitting sigmoidal curves to the percent of remaining protein.  Evidence for the consistency of the \uhalf\ measurements comes from the fact that two independent measurements of the \uhalf\ for our R1-11 parent yielded values that differed by only 0.01 M.  In addition, the \uhalf\ and \thalf\ values are highly correlated (Additional File \ref{add:T50_U50_correlation}), indicating that they provide consistent measures of stability.\pb

For both the \thalf\ and \uhalf\ measurements, the folded P450 was quantified from the carbon monoxide difference spectrum~\cite{Otey2003}.  The microtiter plates containing the P450 samples were first used to read blank spectra at 450 and 490 nm using a Tecan Safire 2 plate reader.  The plates were then incubated for 10 minutes in an airtight oven with carbon monoxide.  The plates were removed form the oven and 10 $\mu$l of 0.1 M sodium hydrosulfite in 1.3 M potassium phosphate (pH 8.0) was immediately added to each well.  After 5-10 minutes, spectra were again read at 450 and 490 nm.  The amount of P450 is proportional to the increase in the signal at 450 nm after this procedure minus the change in the signal at 490 nm.\pb

\ifthenelse{\boolean{publ}}{\end{multicols}}{}

\section*{Mathematical Appendix}
\tableofcontents
\newcounter{appendixnumber}

\subsection*{\refstepcounter{appendixnumber} A.\arabic{appendixnumber} Mathematical background}
\addcontentsline{toc}{section}{A.\arabic{appendixnumber} Mathematical background}
The first purpose of this appendix is to provide mathematical equations that describe the experiments.  The second is to show how four measurements from the experiments can be used to calculate two quantities that describe the topology of the underlying protein neutral network.  We will derive two equations for both quantitites, each in terms of a different measurement.  The fact that the four equations will be seen to yield consistent results provides evidence for the accuracy of the following calculations.\pb

Our calculations are based on a view of neutral protein evolution as a process constrained by a stability threshold, a view that we originally introduced to explain experimental protein mutagenesis results~\cite{Bloom2005}.  The calculations closely parallel our earlier work~\cite{Bloom2007}, which is in turn based on a general theoretical treatment of evolution on neutral networks by van Nimwegen and coworkers~\cite{vanNimwegen1999}.  These calculations will probably be most thoroughly understood by first reading those works.  The primary difference between the current calculations and \cite{Bloom2007} is that previously we assumed that the per generation per protein mutation rate $\mu$ was $\ll 1$, so that at each generation a protein was either unmutated (with probability $1-\mu$) or experienced a single mutation (with probability $\mu$).  In contrast, here we allow the mutation rate to be arbitrarily large, so that a protein may experience multiple mutations in a single generation (in this sense the calculations resemble the generalization by Wilke~\cite{Wilke2001} of \cite{vanNimwegen1999}).  Specifically, let $f_m$ be the probability that a protein experiences $m$ mutations in a single generation.  Here we derive results for arbitrary $f_m$, and then approximations relevant to the form of $f_m$ in the experiments.  In the limiting case of small mutation rate (where $f_0 = 1-\mu$, $f_1 = \mu$, and $f_m = 0$ for $m > 1$), the calculations here reduce to those in \cite{Bloom2007}.  Proteins evolving in nature typically experience very low mutation rates, so \cite{Bloom2007} probably offers the best description of natural protein evolution.  The calculations presented here are designed to specifically treat the evolutionary dynamics of the experiments.\pb

A protein's thermodynamic stability is described by its free energy of folding, \dgf, with more negative values indicating more stable proteins.  As described in several previous papers~\cite{Bloom2005,Bloom2007,Bloom2006}, we assume that selection requires a protein to fold with some minimal stability \dgfmin, so that a protein adequately folds if and only if $\dgf \le \dgfmin$.  The amount of extra stability a protein possesses relative to the stability threshold is given by $\dgfextra = \dgf - \dgfmin$; note that all folded proteins will have $\dgfextra \le 0$.  We further assume that as long as $\dgfextra \le 0$, selection is indifferent to the exact amount of extra stability that a protein possesses (see \cite{Bloom2007} for a discussion of the limitations of this assumption).  We conceptually divide the continuous variable of protein stability into small discrete bins of width $b$.  Specifically, a protein is in bin $i$ if it has \dgfextra\ between $\left(1-i\right)b$ and $-ib$, where $i = 1, 2, \ldots$.  Mutating a protein changes its stability by an amount \ddg\ (defined as the stability of the mutant protein minus the stability of the initial protein), and so may move it to a new stability bin.  In \cite{Bloom2007}, we defined a matrix \W\ with elements \Wij\ giving the transition probabilities that a single mutation changes a protein's stability from bin $j$ to bin $i$.  We noted that \W\ could be computed from the distribution of \ddg\ values for all single mutations, and argued that \W\ remains fairly constant during neutral evolution since the distribution of \ddg\ values remains relatively unchanged.  However, we emphasize that (as discussed in detail in \cite{Bloom2007}) the constancy of the \ddg\ distribution remains an assumption, albeit one that has now been shown to be quite accurate for lattice proteins~\cite{Bloom2005,Bloom2007,Wilke2005} and provide a consistent theoretical explanation for a growing body of experimental results (the current work as well as \cite{Bloom2005}).   \pb

Since we are allowing for larger mutation rates, and we must consider the possibility that a protein's stability might change due to multiple mutations at a single generation.  Therefore, we make a more general definition of \Wijm\ as the probability that $m$ random mutations to a protein in stability bin $j$ change its stability to bin $i$, and let \Wm\ be the matrix with elements \Wijm.  Note that \Wm\ only describes mutations that cause transitions from one folded protein to another, since the stability bins $i = 1, 2, \ldots$ all correspond to folded proteins.  As before~\cite{Bloom2007}, we assume that \Wm\ is roughly constant during evolution, meaning that the distribution of \ddg\ values for multiple mutations is roughly constant during neutral evolution.  Note that if $m = 1$, then \Wm\ is just the matrix \W\ that can be computed from the distribution of single-mutant \ddg\ values~\cite{Bloom2007}.  We will now use the matrices \Wm\ to calculate the following characteristics of a population that has evolved to equilibrium: the distribution of stabilities, the average number of mutations \mavg\ accumulated after $T$ generations, and the average fraction \avgF\ of stably folded proteins in the population.  We then introduce a few approximations (that should be quite accurate for the experimental work in this paper) that greatly simplify these calculations.  Finally, we relate the calculations to properties of the underlying protein neutral network.\pb

As described generally by van Nimwegen and coworkers~\cite{vanNimwegen1999}, the evolutionary dynamics depend on whether the evolving population tends to be monomorphic or highly polymorphic.  When the per sequence per generation mutation rate $\mu$ is $\ll 1$, whether the population is mostly monomorphic or highly polymorphic is determined by the product of the population size $\Npop$ and $\mu$: when $\Npop\mu \ll 1$ the population is mostly monomorphic, and when $\Npop\mu \gg 1$ the population is highly polymorphic.  However, with multiple mutations per generation, $\Npop\mu$ is no longer an appropriate parameter to distinguish between mono- and polymorphism, since if the population size is sufficiently small the population can still be monomorphic even if there are multiple mutations per generation.  Specifically, in one set of experiments we constrained the population to be monomorphic (by maintaining a population size of one), but still allowed the single protein in this population to experience more than one mutation at a generation.  So we instead denote the populations as either monomorphic or polymorphic.  We indicate quantities calculated for the monomorphic population by the subscript $M$ (i.e. \avgFmono) and those calculated for the polymorphic population by the subscript $P$ (i.e. \avgFpoly).\pb

\subsection*{\refstepcounter{appendixnumber} A.\arabic{appendixnumber} Monomorphic limit}
\addcontentsline{toc}{section}{A.\arabic{appendixnumber} Monomorphic limit}
In the limit of a completely monomorphic population, all of the proteins are in a single stability bin.  Let $p_i\left(t\right)$ be the probability that the population is in stability bin $i$ at time $t$, and let $\p\left(t\right)$ be the column vector with elements $p_i\left(t\right)$.  At each generation there is a probability $f_0$ that there is no mutation that becomes fixed in the population, a probability of $\sum\limits_{m=1}^{\infty}f_m\sum\limits_{j}\Wijm p_j$ that the population experiences a mutational event (which could be a single mutation or several simultaneous mutations) that moves it into bin $i$, and a probability $\sum\limits_{m=1}^{\infty}f_m p_i\sum\limits_{j}\Wjim$ that the population is in bin $i$ and experiences one or more mutations that move it to another bin of stably folded proteins.  Define $\nuim = \sum\limits_{j}\Wjim$ to be the fraction of $m$-mutants of a protein in bin $i$ that still fold, and let \Vm\ be the matrix with diagonal elements given by $\Viim = \nuim$ and all other elements zero. The time evolution of \p\ is 
\begin{equation}
\label{eq:ptmono}
\p\left(t+1\right) = \left[\I + \sum\limits_{m=1}^{\infty}f_m\left(\Wm - \Vm\right)\right]\p\left(t\right)
\end{equation}
where \I\ is the identity matrix.  Note that mutations that destabilize a protein beyond the stability threshold are immediately lost to natural selection, and so leave the population in its original stability bin.  This describes the experiments for the monomorphic populations, where we retain the parental sequence if the single mutant we generate is nonfunctional.  Equation \ref{eq:ptmono} corresponds to Equation (1) of \cite{Bloom2007}, and the blind ant random walk described by van Nimwegen and coworkers~\cite{vanNimwegen1999}. \pb

Equation \ref{eq:ptmono} describes a Markov process with a non-negative, irreducible, and acyclic transition matrix, and so \p\ approaches a unique stationary distribution (equilibrium value) of \pmono\ given by the eigenvector equation
\begin{equation}
\label{eq:pmono}
\pmono = \left[\I + \sum\limits_{m=1}^{\infty}f_m\left(\Wm - \Vm\right)\right]\pmono.
\end{equation}
Once \p\ has reached equilibrium, the average fraction of proteins that still stably fold at each generation is
\begin{equation}
\label{eq:avgFmono}
\avgFmono = \e\left(f_0\I + \sum\limits_{m=1}^{\infty} f_m\Wm\right)\pmono 
\end{equation}
where $\e = \left(1, \ldots, 1\right)$ is the unit row vector.\pb

To calculate \mavgmono, the average number of mutations accumulated after $T$ generations once the population has equilibrated, we note that at each generation there is a probability of $f_m p_j \sum\limits_i\Wijm$ that a randomly chosen protein is in bin $j$, experiences $m$ mutations, and still stably folds.  The average number of mutations accumulated in a single generation is simply the average of $m$ weighted over this probability.  So summing over all values of $m$ and $j$, we see that
\begin{equation}
\label{eq:mavgmono}
\mavgmono = T\e\sum\limits_{m=0}^{\infty}m f_m \Wm \pmono.
\end{equation}
This equation corresponds to Equation (6) of \cite{Bloom2007}, which was derived using an embedded Markov process formalism.  Here we have foregone this formalism for the more intuitive argument presented above, since we do not attempt to calculate higher moments of the number of mutations.  \pb

\subsection*{\refstepcounter{appendixnumber} A.\arabic{appendixnumber} Polymorphic limit}
\addcontentsline{toc}{section}{A.\arabic{appendixnumber} Polymorphic limit}
In the limit when the population is highly polymorphic, at each generation there are sequences in many different stability bins.  In this case, we describe the distribution of stabilities by the column vector $\x\left(t\right)$, with element $x_i\left(t\right)$ giving the fraction of proteins in stability bin $i$ at time $t$.  At generation $t$, the fraction of mutants that continue to fold is
\begin{equation}
\avgFt = \e\left(f_0\I + \sum\limits_{m=1}^{\infty} f_m\Wm\right)\x\left(t\right).
\end{equation}  
Therefore, in order to maintain a constant population size, each remaining protein must produce an average of  $\ratet = \avgFt^{-1}$ offspring. The population therefore evolves according to
\begin{equation}
\label{eq:xtpoly}
\x\left(t+1\right) = \ratet\left(f_0 \I + \sum\limits_{m=1}^{\infty}f_m\Wm\right)\x\left(t\right).
\end{equation}
After the population evolves for a sufficiently long period of time, \x\ will approach an equilibrium value of \xpoly.  At this equilibrium, the average fraction of mutants that fold at each generation is 
\begin{equation}
\label{eq:avgFpoly}
\avgFpoly = \e\left(f_0\I + \sum\limits_{m=1}^{\infty}f_m\Wm\right)\xpoly,
\end{equation}
and the equilibrium reproduction rate is $\rateinf = \avgFpoly^{-1}$.  Therefore,
\begin{equation}
\label{eq:xpoly}
\xpoly = \rateinf\left(f_0 \I + \sum\limits_{m=1}^{\infty}f_m\Wm\right)\xpoly.
\end{equation}
Equations \ref{eq:avgFpoly} and \ref{eq:xpoly} can be combined to show that \xpoly\ and \avgFpoly\ can be calculated from the eigenvector equation
\begin{equation}
\label{eq:eigpoly}
\left(\avgFpoly - f_0\right)\xpoly = \sum\limits_{m=1}^{\infty}f_m\Wm \xpoly,
\end{equation}
with $\left(\avgFpoly-f_0\right)$ the principal eigenvalue of the nonnegative and irreducible matrix $\sum\limits_{m=1}^{\infty}f_m\Wm$.  Equation \ref{eq:eigpoly} corresponds to Equation (14) of \cite{Bloom2007}, Equation (6) of the work by van Nimwegen and coworkers~\cite{vanNimwegen1999}, and Equation (13) of the work by Wilke~\cite{Wilke2001}.\pb

We now calculate \mavgpoly, the average number of mutations accumulated in $T$ generations after the population has equilibrated.  At equilibrium, there is a probability of $f_m x_j \sum\limits_i\Wijm$ that a protein is in bin $j$, experiences $m$ mutations, and still stably folds.  Subsequently, all of these folded proteins produce an average of \rateinf\ offspring. The average number of mutations accumulated in a single generation is simply the average of $m$ weighted over this probability, and then multiplied by the average reproduction rate.  So summing over all values of $m$ and $j$, we obtain
\begin{equation}
\label{eq:mavgpoly}
\mavgpoly = \rateinf T\e\sum\limits_{m=0}^{\infty}m f_m \Wm \xpoly = \frac{T}{\avgFpoly}\e\sum\limits_{m=0}^{\infty}m f_m \Wm \xpoly.
\end{equation}
This equation is the counterpart of Equation (18) of \cite{Bloom2007}, where we have again foregone the embedded Markov process formalism for a more intuitive derivation.\pb

\subsection*{\refstepcounter{appendixnumber} A.\arabic{appendixnumber} Approximations for polymorphic limit}
\addcontentsline{toc}{section}{A.\arabic{appendixnumber} Approximations for polymorphic limit}
We can dramatically simplify the results from the previous sections with several reasonable approximations.  The first approximation is that the \ddg\ values for random mutations are roughly additive, and is supported by a number of experimental studies of the thermodynamic effects of mutations~\cite{Wells1990,Zhang1995,Serrano1993}.  We have previously shown that this approximation can be used to accurately describe experimental protein mutagenesis data with a simple stability threshold model~\cite{Bloom2005}.  Under this approximation, the distribution of net \ddg\ values for multiple mutations can be computed from the distribution of \ddg\ values for single mutations by performing convolutions of the single-mutation \ddg\ distribution~\cite{Bloom2005}, meaning that \Wm\ for arbitrary $m$ can be computed solely from the distribution of \ddg\ values for single mutations.  However, to simplify the equations from previous sections, we need to express \Wm\ for arbitrary $m$ only in terms of \W\ (recall that $\W = \Wone$).  Since \W\ only contains information about stability transitions from folded proteins to other folded proteins, if we make the second approximation that a protein that is destabilized beyond the minimal stability threshold by one mutation is not re-stabilized to a folded protein by a subsequent mutation, then $\Wm = \W^m$.  This approximation that unfolded proteins  are not re-stabilized should be quite accurate since stabilizing mutations tend to be relatively rare and small in magnitude~\cite{Godoy-Ruiz2004,Pakula1989,Matthews1993,Bava2004} (this is the underlying idea behind the Markov chain approximation that was shown to be highly accurate for lattice proteins~\cite{Wilke2005}).  To summarize, if \ddg\ values are roughly additive and stabilizing mutations are rare, we have the approximation 
\begin{equation}
\label{eq:Wapprox}
\Wm \approx \W^m.
\end{equation}\pb

Simplifying the equations of the previous sections also requires assigning a specific functional form to $f_m$, the probability that a sequence undergoes $m$ mutations.  Here we assume that mutations are Poisson distributed among sequences, so that 
\begin{equation}
\label{eq:fm}
f_m = \frac{e^{-\mu}\mu^m}{m!}
\end{equation}
where $\mu = \sum\limits_{m=0}^{\infty}m f_m$ is the average number of mutations per protein per generation.  When the mutations are introduced by error-prone PCR, the Poisson distribution is an excellent approximation to the true theoretical distribution of mutations created by error-prone PCR~\cite{Sun1995,Drummond2005c} provided that $\mu$ is much less than the number of PCR doublings, as is the case in all of the experiments in the current work.\pb

We now use the approximations of Equations \ref{eq:Wapprox} and \ref{eq:fm} to simplify the results given above for the highly polymorphic limit.  We begin by using these approximations to rewrite Equation \ref{eq:eigpoly} as
\begin{eqnarray}
\label{eq:eigpolyapprox1}
\left(\avgFpoly - e^{-\mu}\right)\xpoly &=& e^{-\mu}\sum\limits_{m=1}^{\infty} \frac{\mu^m}{m!} \W^m \xpoly.
\end{eqnarray}
This equation makes clear that \xpoly\ is the principal eigenvector of the matrix $\sum\limits_{m=1}^{\infty}\frac{\mu^m}{m!}\W^m$, therefore \xpoly\ must also be the principal eigenvector of \W.  Now in our earlier work~\cite{Bloom2007}, we defined the principal eigenvector of \W\ as \xinf, called the corresponding eigenvalue \nuinf, and showed that this eigenvalue is shown the average fraction of single mutations that are neutral in a population that is evolving with $\Npop\mu \gg 1$ and $\mu \ll 1$.  Therefore, with the approximation of Equation \ref{eq:Wapprox}, \xpoly\ and \xinf\ are equal, and are both defined by the same eigenvector equation,
\begin{equation}
\label{eq:nuinfeigen}
\nuinf\xpoly = \W\xpoly = \W\xinf = \nuinf\xinf.
\end{equation}
Combining Equations \ref{eq:eigpolyapprox1} and \ref{eq:nuinfeigen} we have,
\begin{eqnarray}
\label{eq:eigpolyapprox2}
\avgFpoly\xpoly 
&=& e^{-\mu}\sum\limits_{m=0}^{\infty}\frac{\left(\mu\nuinf\right)^m}{m!}\xpoly \nonumber \\
&=& e^{-\mu\left(1-\nuinf\right)}\xpoly
\end{eqnarray}
Equation \ref{eq:eigpolyapprox2} can be solved to yield
\begin{equation}
\label{eq:nuinfapprox}
\nuinf = 1 + \frac{\ln{\avgFpoly}}{\mu}.
\end{equation}
Similarly, we can simplify Equation \ref{eq:mavgpoly},
\begin{eqnarray}
\label{eq:mavgpolyapprox}
\mavgpoly &=& \frac{T}{\avgFpoly}\e\sum\limits_{m=1}^{\infty}m f_m \Wm \xpoly \nonumber \\
&=&T e^{\mu\left(1 - \nuinf\right)}\sum\limits_{m=1}^{\infty}me^{-\mu}\frac{\mu^m}{m!}\e\W^m\xpoly \nonumber \\
&=& T e^{-\mu\nuinf} \sum\limits_{m=1}^{\infty}m\frac{\left(\mu\nuinf\right)^m}{m!} \nonumber \\
&=& T\mu\nuinf e^{-\mu\nuinf} \sum\limits_{m=0}^{\infty}\frac{\left(\mu\nuinf\right)^m}{m!} \nonumber \\
&=&T\mu\nuinf.
\end{eqnarray}
Solving this equation for \nuinf\ yields
\begin{equation}
\label{eq:nuinfapprox2}
\nuinf = \frac{\mavgpoly}{T\mu}.
\end{equation}\pb

\subsection*{\refstepcounter{appendixnumber} A.\arabic{appendixnumber} Approximations for monomorphic limit}
\addcontentsline{toc}{section}{A.\arabic{appendixnumber} Approximations for monomorphic limit}
We now simplify the equations for the monomorphic limit.  This requires several further approximations.  We begin by approximating that the stability probability distribution \pmono\ given by Equation \ref{eq:pmono} by the distribution \psmall\ defined in \cite{Bloom2007} as satisfying 
\begin{equation}
\label{eq:psmall}
0 = \left(\W - \V\right)\psmall.
\end{equation}
The basic rationale behind approximating \pmono\ with \psmall\ is that Equation \ref{eq:pmono} can be viewed as a perturbation to Equation \ref{eq:psmall}~\cite{Franklin1968}.  Essentially, \psmall\ is an eigenvector of the matrix $\W - \V$ while \pmono\ is the corresponding eigenvector of the matrix $\W - \V + \sum\limits_{m=2}^{\infty}\frac{\mu^{m-1}}{m!}\left(\W^m - \Vm\right)$.  The latter matrix can be viewed as a perturbation to the first, since the sum $\sum\limits_{m=2}^{\infty}\frac{\mu^{m-1}}{m!}\left(\W^m - \Vm\right)$ is small.  This smallness is due to the fact that $\W^m$ tends to zero with large $m$, causing $\Vm$ to tend towards the identity matrix.  In addition, the $\mu^m/m!$ terms tend to zero with large $m$.  Therefore, the terms in the summation are all simply either a perturbation to $\W - \V$ or involve subtracting terms that are fractions of the identity matrix.  The perturbations lead to bounded changes in the eigenvectors~\cite{Franklin1968}, while the identity matrix terms do not change the eigenvectors.  Below we give a more rigorous justification of the assumption that \pmono\ is approximately equal to \psmall.\pb

We need one additional approximation to make further progress.  Both Equations \ref{eq:avgFmono} and \ref{eq:mavgmono} contain terms of the form $\Wm\psmall$, and even if we use Equation \ref{eq:Wapprox} to rewrite these terms as $\W^m\psmall$, there are no further clear simplifications.  However, any probability vector that is multiplied repeatedly by \W\ and normalized will eventually converge to $\xinf = \xpoly$ (since this is the principal eigenvector of \W).  We make the approximation that this convergence is sufficiently rapid to be essentially complete after a single multiplication.  This approximation is supported by both protein mutagenesis studies~\cite{Bloom2005,Shafikhani1997,Guo2004} that indicate that proteins rapidly converge to an exponential decline in the fraction folded (indicating the stability distribution has equilibrated, as discussed below, and by lattice protein studies showing the same~\cite{Bloom2005,Wilke2005}.   Therefore, we make the approximation that $\e\W^m\psmall = \nusmall\e\W^{m-1}\xinf = \nusmall\nuinf^{m-1}$ where $\nusmall = \e\W\psmall$ has the same definition as in \cite{Bloom2007}, where it was defined as the average fraction of functional single mutants of a population evolving with $\mu \ll 1$ and $N\mu \ll 1$.\pb

We use these approximations to simplify Equation \ref{eq:avgFmono} as
\begin{eqnarray}
\label{eq:avgFmonoapprox}
\avgFmono
&=& \e\left(f_0\I + \sum\limits_{m=1}^{\infty} f_m\Wm\right)\pmono \nonumber \\
&=& e^{-\mu}\left[1 + \sum\limits_{m=1}^{\infty}\frac{\mu^m}{m!}\e\W^m\psmall\right] \nonumber \\
&=& e^{-\mu}\left[1 + \mu\nusmall\sum\limits_{m=1}^{\infty}\frac{\left(\mu\nuinf\right)^{m-1}}{m!}\right]\nonumber \\
&=& e^{-\mu}\left[1 + \frac{\nusmall}{\nuinf}\left(-1 + \sum\limits_{m=0}^{\infty}\frac{\left(\mu\nuinf\right)^m}{m!}\right)\right] \nonumber \\
&=& e^{-\mu}\left[1 + \frac{\nusmall}{\nuinf}\left(e^{\mu\nuinf} - 1\right)\right].
\end{eqnarray}
Solving this equation for \nusmall, we find
\begin{equation}
\label{eq:nusmallapprox}
\nusmall = \frac{\nuinf\left(\avgFmono\e^{\mu} - 1\right)}{e^{\mu\nuinf} - 1}.
\end{equation}
We now use the approximations to simplify Equation \ref{eq:mavgmono} as
\begin{eqnarray}
\label{eq:mavgmonoapprox}
\mavgmono 
&=& T\e\sum\limits_{m=0}^{\infty}m f_m \Wm \pmono \nonumber \\
&=& Te^{-\mu}\sum\limits_{m=1}^{\infty}m\frac{\mu^m}{m!}\e\W^m\psmall \nonumber \\
&=& Te^{-\mu}\nusmall\sum\limits_{m=1}^{\infty}m\frac{\mu^m}{m!}\nuinf^{m-1} \nonumber \\
&=& \mu Te^{-\mu}\nusmall\sum\limits_{m=0}^{\infty}\frac{\left(\mu\nuinf\right)^m}{m!} \nonumber \\
&=& \mu T\nusmall e^{\mu\left(\nuinf-1\right)}.
\end{eqnarray}
Solving this equation for \nusmall\ yields
\begin{equation}
\label{eq:nusmallapprox2}
\nusmall = \frac{\mavgmono e^{\mu\left(1 - \nuinf\right)}}{\mu T}.
\end{equation}\pb

To recap, we now have equations to calculate \nuinf\ and \nusmall\ from experimentally measurable quantities.  Equations \ref{eq:nuinfapprox} and \ref{eq:nuinfapprox2} allow us to calculate \nuinf\ from \avgFpoly\ and \mavgpoly, respectively.  Given this calculated value of \nuinf, Equations \ref{eq:nusmallapprox} and \ref{eq:nusmallapprox2} then allow us to calculate \nusmall\ from \avgFmono\ and \mavgmono, respectively.  The fact that we have two equations each for \nuinf\ and \nusmall\ allows us to assess the self-consistency of the approach.\pb

\subsection*{\refstepcounter{appendixnumber} A.\arabic{appendixnumber} Interpretation in terms of neutral networks}
\addcontentsline{toc}{section}{A.\arabic{appendixnumber} Interpretation in terms of neutral networks}
Throughout the preceding calculations, we have referred to \nuinf\ and \nusmall\ as we defined them in \cite{Bloom2007}: namely, as the average neutrality of protein populations evolving with $\mu \ll 1$ and $N\mu$ either $\gg 1$ or $\ll 1$, respectively.  However, van Nimwegen and coworkers~\cite{vanNimwegen1999} have shown that they can also be interpreted in terms of the underlying neutral network.  In the experiments we make mutations at the nucleotide (rather than amino acid) level, so each point in our sequence space corresponds to a different gene.  Every gene that yields an amount of protein sufficient to hydroxylate the twelfth carbon of 12-\textit{p}-nitrophenoxydodecanoic acid with at least 75\% of the total activity conferred by the original R1-11 parent gene represents a node on this neutral network.  We note that in the experiments (and also usually in natural evolution), the edges on the neutral network are not all completely equivalent or fully undirected, since some mutations are more likely to occur than others (for example, error-prone PCR with \textit{Taq} polymerase is more likely to cause an A$\rightarrow$G mutation than an A$\rightarrow$C mutation).  In the analysis that follows, we ignore this complication and assume all neutral network edges are equivalent.  \pb

In an extremely insightful analysis, van Nimwegen and coworkers~\cite{vanNimwegen1999} have shown that important characteristics of a neutral network can be inferred from evolutionary quantities.  Specifically, they have shown that if a population is evolving with $\mu \ll 1$ and $\Npop\mu \gg 1$, then the average neutrality (which we have denoted by \nuinf) is equal to the principal eigenvalue of the adjacency matrix of the neutral network, normalized by the network coordination number (number of possible connections per node).   In addition, they pointed out that a population evolving with $\mu \ll 1$ and $\Npop\mu \ll 1$ moves like a blind ant random walk, meaning that the average neutrality (which we have denoted by \nusmall) is equal to the average connectivity of a neutral network node divided by the network coordination number.  In our P450 experiments, we have measured the values needed to estimate \nuinf\ and \nusmall\ using Equations \ref{eq:nuinfapprox}, \ref{eq:nuinfapprox2}, \ref{eq:nusmallapprox}, and \ref{eq:nusmallapprox2}.  Using the final values listed in Table \ref{tab:evolutionsummary}, $\avgFpoly = 0.50$ and $\avgFmono = 0.39$.  Taking the final nucleotide mutation values from Table \ref{tab:evolutionsummary}, $\mavgpoly / T = 0.69$ and $\mavgmono / T = 0.31$.  The average mutation rate, computed from the unselected population, is $\mu = 1.40$.  So using Equation \ref{eq:nuinfapprox}, $\nuinf = 0.53$, while using Equation \ref{eq:nuinfapprox2}, $\nuinf = 0.49$.  The consistency of these two values supports the idea that the calculations above accurately describe the evolutionary process.  Taking the average value of these two measurement as $\nuinf = 0.51$, we can then use Equations \ref{eq:nusmallapprox} and \ref{eq:nusmallapprox2} to calculate \nusmall.  We calculate values of 0.28 and 0.43, respectively.  These estimates differ by more than those for \nuinf, perhaps because additional approximations have gone into the derivation of the relevant equations (in addition, we have made no attempt to carry out the rather complex propagation of the sampling errors of Table \ref{tab:evolutionsummary}).  However, the values are still in a similar range.  Taking the average of these two values, we estimate that $\nusmall = 0.35$.  So overall, we predict that each functional P450 gene should have an average fraction of 0.35 of its sequence nearest neighbors also encoding a functional gene, for an average of about 1,500 neighbor genes.  We predict that the principal eigenvalue of the neutral network adjacency matrix is 0.51 $\times 3L$.  The fact that principal eigenvalue exceeds the average connectivity indicates that the neutral network is not a regular graph, but instead has some nodes more highly connected than others.\pb

The value for \nuinf\ calculated above can also be related to measurements from protein mutagenesis experiments.  A number of studies~\cite{Bloom2005,Shafikhani1997,Guo2004} have observed that the probability that a protein remains functional after $m$ mutations falls off exponentially with the number of mutations.  In fact, the decline is not always exponential for the first few mutations if the starting protein has especially high or low stability~\cite{Bloom2005} or activity~\cite{Bershtein2006}, but will still converge to this exponential form after a few mutations~\cite{Bloom2005,Wilke2005,Bloom2007b}.  The stability threshold model can be used to relate this decline to \nuinf, as is done indirectly in the Markov chain approximation of \cite{Wilke2005}.  Here we make that connection explicit.  The initial protein has a stability that falls into some stability bin $i$.  Therefore, its stability can be described by the column vector \yzero, which has element $i$ equal to one and all other elements equal to zero.  Now imagine constructing all single mutants of this protein.  The fraction of these single mutants that still fold is just $\e\W\yzero$, and the distribution of stabilities among the single mutants is $\yone = \W\yzero$ (note that the elements of \yone\ no longer sum to one).  Similarly, after $m$ mutations, the fraction of mutants that still fold is $\e\Wm\yzero$, and the distribution of stabilities among the $m$-mutants is $\ym = \Wm\yzero$.  With the approximation of Equation \ref{eq:Wapprox}, $\ym = \W^m\yzero$.  This makes it clear that \ym\ will converge to a vector proportional to \xinf, the principal eigenvector of \W.  Once this convergence is complete, each new mutation simply reduces the fraction of mutants that fold by a factor of \nuinf, the principal eigenvalue of \W\ (and the spectral radius of the neutral network normalized by the coordination number).  Therefore, what we have called \nuinf\ in the present work and \cite{Bloom2007} is equal to what is called $x$ in \cite{Guo2004}, $q$ in \cite{Shafikhani1997}, and $\langle \nu \rangle$ in \cite{Bloom2005}.  The major difficulty that is faced in extracting \nuinf\ by the method of those three studies~\cite{Bloom2005,Shafikhani1997,Guo2004} is that it is not possible to directly assay mutants with $m$ mutations, but instead only possible to assay a set of mutants with a distribution of $m$.  All three studies use different (and valid) methods to account for this distribution, but this accounting is still difficult because most of the functional mutants come from the low $m$ end of the distribution.  This makes it hard to get accurate values for the fraction functional after large numbers of mutations, since most of the functional mutants in the set come from sequences with few mutations.  For this reason, we believe the current method of measuring \nuinf\ is more accurate.  A second caution about comparing values of \nuinf\ from different studies is that its value depends on the nucleotide error-spectrum of the experiment, since different mutagenesis methods create different distributions of nucleotide and amino acid mutation types.  \pb

We also briefly mention how we arrived at an estimate of \nuinf\ for 3-methyladenine DNA glycosylase from the data of \cite{Guo2004}.  This paper reports that a fraction $x = 0.34$ of amino acid mutations inactivate the protein.  We would like to determine the fraction \nuinf\ of nucleotide mutations that do not inactivate the protein.  Roughly 75\% of random mutations to a gene will be synonymous.  Therefore, $m$ amino acid mutations should cause about $4m/3$ nucleotide mutations.  The study of \cite{Guo2004} measures that after $m$ mutations, a fraction $\left(1-x\right)^m$ of the mutants are functional.  That means that $\nuinf^{4m/3}$ fraction should be functional.  Equating these expressions yields $\nuinf = \exp\left(\frac{3}{4}\log\left(1-x\right)\right)$.  So using $x = 0.34$, we arrive at $\nuinf = 0.73$. \pb

\subsection*{\refstepcounter{appendixnumber} A.\arabic{appendixnumber} Detailed justification for approximating \pmono\ by \psmall}
\addcontentsline{toc}{section}{A.\arabic{appendixnumber} Detailed justification for approximating \pmono\ by \psmall}
Here we provide a detailed justification for the approximation that \pmono\ is about equal to \psmall.  In the monomorphic limit, the time evolution of \p\ is given by Equation \ref{eq:ptmono}, and the stationary distribution \pmono\ is given by Equation \ref{eq:pmono}.  We assume the approximations of Equations \ref{eq:Wapprox} and \ref{eq:fm} and show that we can approximate \pmono\ by \psmall, where \psmall\ is given by Equation \ref{eq:psmall}.  To justify this approximation, we insert \psmall\ into the right hand side of Equation \ref{eq:ptmono} and ask to what extent \psmall\ is left unaltered by the dynamics. If \psmall\ is found to be stationary to good approximation then, by uniqueness of the stationary distribution of an ergodic process, \psmall\ would be a good approximation to \pmono. \pb

We therefore suppose that at some time $t$ the distribution is given by \psmall\ and compute, using Equation \ref{eq:ptmono}, the change in \psmall\ after one generation. The new distribution at time $t+1$ is given by
\begin{equation}
\label{eq:ptmono2}
\p\left(t+1\right) = \left[\I + \sum\limits_{m=1}^{\infty}f_m\left(\W^m - \Vm\right)\right]\psmall.
\end{equation}
Using $\left(\V - \W\right)\psmall = 0$, and taking components of the above equation, we obtain
\begin{equation}
\label{eq:ptmono3}
p_i\left(t+1\right) = p_{0i}+\sum\limits_{m=2}^{\infty}f_m\left[\left(\W^m - \Vm\right)\psmall\right]_i.
\end{equation}
Thus \psmall\ would be an approximately stationary distribution of the dynamics if $\lvert \sum\limits_{m=2}^{\infty}f_m\left[\left(\W^m - \Vm\right)\psmall\right]_i\rvert \ll p_{0i}$. We now proceed to show that this will be the case in most situations of interest by deriving upper and lower bounds on the second term of the right hand side of Equation \ref{eq:ptmono3}.\pb

Consider first the term $\left(\W^m\psmall\right)_i$, which can be written as
\begin{align}
\label{eq:wm1}
\left(\W^m\psmall\right)_i &= \sum_{k_1,\ldots,k_m} W_{ik_1}W_{k_1 k_2} \cdots W_{k_{m-1} k_m}p_{0k_m} \nonumber \\
&= \sum_{k_1,\ldots,k_{m-1}} W_{ik_1}W_{k_1 k_2} \cdots W_{k_{m-2} k_{m-1}}\nu_{k_{m-1}}p_{0k_{m-1}},
\end{align} 
where we have used $\W\psmall = \V\psmall$ in the second equality. We now note that $\nu_k \leq \nu_{\rm max}$ for all $k$, where $\nu_{\rm max}$ is the maximum neutrality, maximized over all bins. This leads to the successive inequalities
\begin{align}
\label{eq:wm2}
\left(\W^m\psmall\right)_i &\leq \nu_{\rm max}\sum_{k_1,\ldots,k_{m-1}} W_{ik_1}W_{k_1 k_2} \cdots W_{k_{m-2} k_{m-1}}p_{0k_{m-1}} \nonumber \\
&= \nu_{\rm max}\sum_{k_1,\ldots,k_{m-2}} W_{ik_1}W_{k_1 k_2} \cdots W_{k_{m-3} k_{m-2}}\nu_{k_{m-2}}p_{0k_{m-2}} \nonumber \\
&\leq  \nu_{\rm max}^2\sum_{k_1,\ldots,k_{m-2}} W_{ik_1}W_{k_1 k_2} \cdots W_{k_{m-3} k_{m-2}}p_{0k_{m-2}} \nonumber \\
&\leq \nu_{\rm max}^{m-1} \sum_{k_1}W_{ik_1}p_{0k_1},
\end{align}
yielding the upper bound
\begin{equation}
\label{eq:ineq1}
\left(\W^m\psmall\right)_i \leq \nu_{\rm max}^{m-1} \nu_i p_{0i}.
\end{equation}
In an identical manner, we obtain the lower bound
\begin{equation}
\label{eq:ineq2}
\left(\W^m\psmall\right)_i \geq \nu_{\rm min}^{m-1} \nu_i p_{0i},
\end{equation}
where $\nu_{\rm min}$ is the smallest neutrality, minimized over all bins. Note that both inequalities above become exact equalities when all bins have the same neutrality $\nu$, which could be interpreted as either $\nu_{\rm min}$ or $\nu_{\rm max}$. \pb

Having obtained inequality constraints on $\left(\W^m\psmall\right)_i$, we now consider the term $\left(\Vm\psmall\right)_i$, which can be written as
\begin{align}
\label{eq:vm1}
\left(\Vm\psmall\right)_i &= p_{0i}\nu_{i,m} \nonumber \\
&= p_{0i} \sum_j (\W^m)_{ji} \nonumber \\
&= p_{0i}\sum_{j,k_1,\ldots,k_{m-1}} W_{jk_1}W_{k_1 k_2}\cdots W_{k_{m-1}i}\nonumber \\
&= p_{0i}\sum_{k_1,\ldots,k_{m-1}} \nu_{k_1}W_{k_1 k_2}\cdots W_{k_{m-1}i}\nonumber \\
&\leq p_{0i}\nu_{\rm max}\sum_{k_1,\ldots,k_{m-1}} W_{k_1 k_2}\cdots W_{k_{m-1}i}\nonumber \\
&\leq p_{0i}\nu_{\rm max}^{m-1}\sum_{k_{m-1}}W_{k_{m-1}i},
\end{align}
which yields an identical upper bound to that on $\left(\W^m\psmall\right)_i$, namely,
\begin{equation}
\label{eq:ineq3}
\left(\V^m\psmall\right)_i \leq \nu_{\rm max}^{m-1} \nu_i p_{0i},
\end{equation}
and similarly
\begin{equation}
\label{eq:ineq4}
\left(\V^m\psmall\right)_i \geq \nu_{\rm min}^{m-1} \nu_i p_{0i}.
\end{equation}
It should again be noted that both the above inequalities become exact equalities when all bins have a common neutrality $\nu$.\pb

We are now in a position to estimate bounds on the magnitude of the second term of Equation \ref{eq:ptmono3}. Using the four inequalities of 
Equations \ref{eq:ineq1}, \ref{eq:ineq2}, \ref{eq:ineq3}, and \ref{eq:ineq4} above, we have
\begin{equation}
-\left(\nu_{\rm max}^{m-1}-\nu_{\rm min}^{m-1}\right)\nu_i p_{0i}\leq \left[\left(\W^m - \Vm\right)\psmall\right]_i \leq \left(\nu_{\rm max}^{m-1}-\nu_{\rm min}^{m-1}\right)\nu_i p_{0i},
\end{equation}
or equivalently,
\begin{equation}
\label{eq:fundineq}
\left\lvert \left[\left(\W^m - \Vm\right)\psmall\right]_i \right\rvert \leq \left(\nu_{\rm max}^{m-1}-\nu_{\rm min}^{m-1}\right)\nu_i p_{0i},
\end{equation}
where the inequality above becomes an exact equality when all bins have the same neutrality. However, in this limit, the right hand side of the above equation vanishes, and therefore the second term of Equation \ref{eq:ptmono3} is identically zero in this case, giving the result that \pmono\ is exactly equal to \psmall\ when all bins have the same neutrality, even if $\mu$ is  arbitrarily large.\pb

We now carry out the sum over $m$ to obtain an upper bound on the second term of Equation \ref{eq:ptmono3} in the more general and realistic case of unequal neutrality bins. Using Equation \ref{eq:fundineq} and the specific Poisson form of $f_m$, we obtain an upper bound on the fractional change in $p_{0i}$ in one generation:
\begin{align}
\left\lvert \frac{p_i(t+1)-p_{0i}}{p_{0i}}\right\rvert &\leq \nu_i e^{-\mu}  \sum_{m=2}^{\infty}\frac{\mu^m}{m!}\left(\nu_{\rm max}^{m-1}-\nu_{\rm min}^{m-1}\right) \nonumber \\
&= \nu_i e^{-\mu}\left[\frac{e^{\mu \nu_{\rm max}}-1}{\nu_{\rm max}} - \frac{e^{\mu \nu_{\rm min}}-1}{\nu_{\rm min}}\right].
\end{align}
The above bound vanishes for small $\mu$, is an increasing function of $\nu_{\rm max} - \nu_{\rm min}$, and is typically much smaller than 1. An extreme estimate of the size of the fractional change can be made when $\nu_{\rm max}=1$ and $\nu_{\rm min}=0$.  In this case, using $\mu=1.4$ (the value in our experiments), the above inequality simplifies to
\begin{equation}
\left\lvert \frac{p_i(t+1)-p_{0i}}{p_{0i}}\right\rvert \leq \nu_i \left(1-e^{-\mu}-\mu e^{-\mu}\right) \simeq 0.41 \nu_i.
\end{equation}
Noting that $\nu_i <1$, the fractional change in $p_{0i}$ is therefore reasonably controlled even in the most extreme case. For realistic situations, the fractional change in $p_{0i}$ is expected to be much lower, thus justifying the use of \psmall\ as the stationary distribution of the dynamics of Equation \ref{eq:ptmono}.\pb

\newpage

\section*{Author Contributions}
JDB and FHA designed the project and wrote the paper.  JDB and ZL performed the bulk of the experiments; OSV assisted with the experiments.  JDB and DC analyzed the data.  JDB and AR performed the theoretical work.

\section*{Acknowledgments}
\ifthenelse{\boolean{publ}}{\small}{}
We thank Claus O Wilke for helpful advice and comments.  JDB is supported by a HHMI predoctoral fellowship.  ZL and DC were supported by Summer Undergraduate Research Fellowships from the California Institute of Technology.  AR is supported by NSF grants CCF 0523643 and FIBR 0527023.

{\ifthenelse{\boolean{publ}}{\footnotesize}{\small}
 \bibliographystyle{bmc_article}  
   \bibliography{/Users/bloom/references/references} }

\newpage

\section*{Figures}
\newcounter{figurenumber}

\centerline{\includegraphics[width=2.3in]{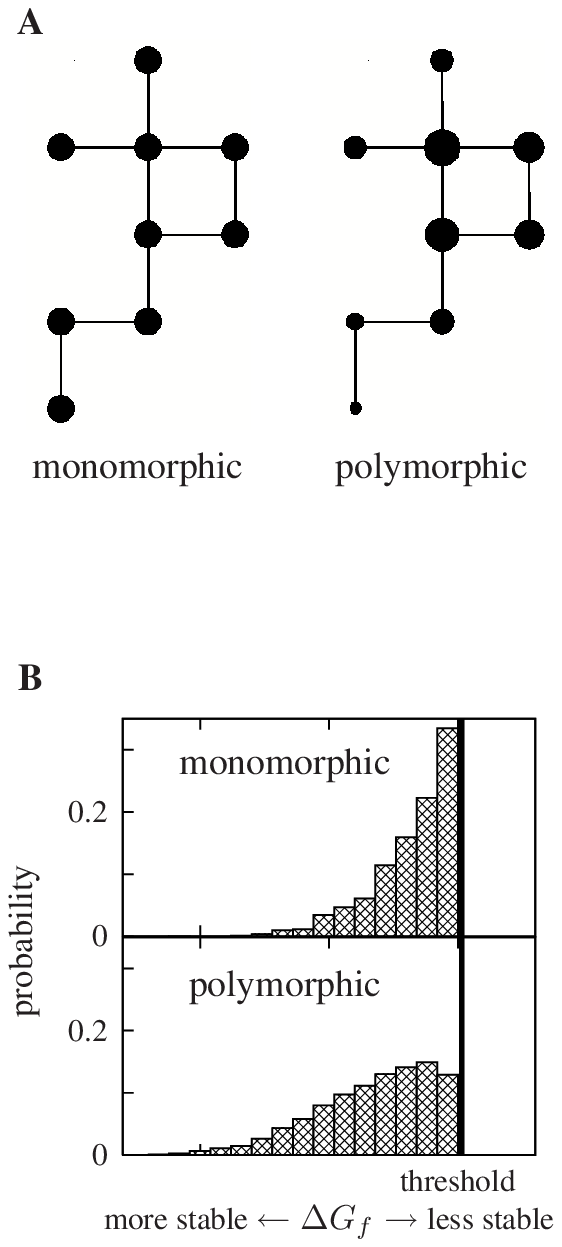}}
\subsection*{\refstepcounter{figurenumber}\label{fig:model}Figure \arabic{figurenumber} - Theoretical views of the evolution of protein mutational robustness and stability.}
{\bf (A)} Theory predicts that a mostly monomorphic population is equally likely to occupy any node of its neutral network, while a highly polymorphic population will prefer more connected nodes~\cite{vanNimwegen1999}.  Node sizes are drawn proportional to the occupation probabilities.  {\bf (B)} Proteins evolving in a highly polymorphic population are predicted to be more stable than their counterparts in a mostly monomorphic population~\cite{Bloom2007}.  The histograms illustrate the distributions of stabilities for the two cases.  The increased stability is a biophysical manifestation of excess mutational robustness, since more stable proteins are more mutationally robust~\cite{Bloom2005,Besenmatter2007,Bloom2006}.
\newpage

\centerline{\includegraphics[width=6.5in]{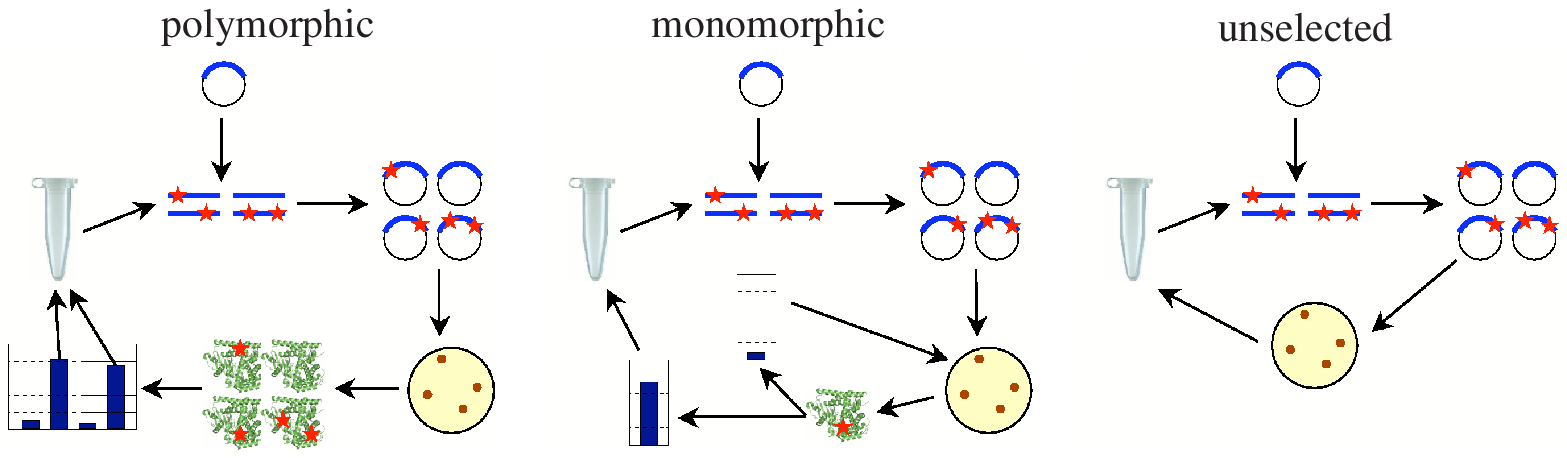}}
\subsection*{\refstepcounter{figurenumber}\label{fig:experiment}Figure \arabic{figurenumber} - Outline of the neutral evolution experimental procedure.}  
For the {\bf polymorphic} population, error-prone PCR was used to generate mutant P450 genes.  These genes were ligated into a plasmid and transformed into \textit{E. coli}.  Individual mutants (435) were picked, expressed in \textit{E. coli}, and assayed for enzymatic activity.  All mutants that met the selection criterion contributed an equal amount of plasmid DNA as template for the next generation of error-prone PCR.  The {\bf monomorphic} populations were treated similarly, except only a single mutant was assayed at each generation.  If this mutant met the selection criterion then it became the template for the next generation of error-prone PCR; otherwise at the next generation another colony was picked from the same plate.  In the {\bf unselected} populations a single mutant was picked and used as the template for the next generation of error-prone PCR.
\newpage

\centerline{\includegraphics[width=2.5in]{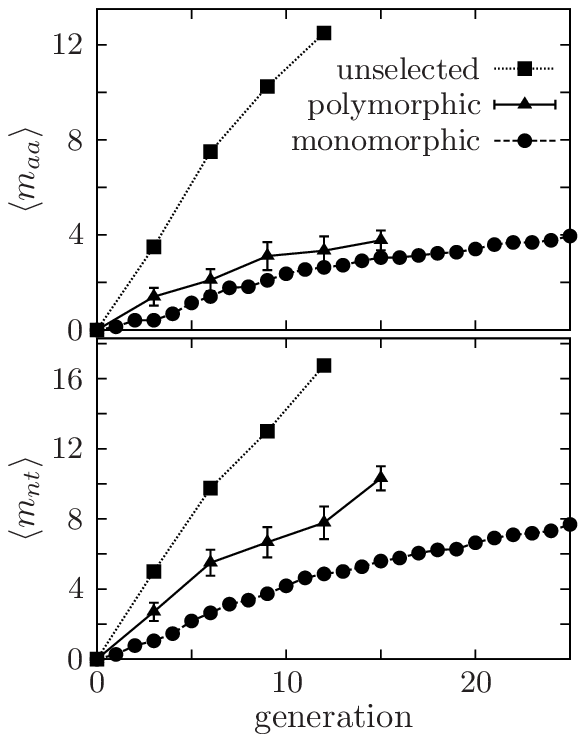}}
\subsection*{\refstepcounter{figurenumber}\label{fig:mutations}Figure \arabic{figurenumber} - Accumulation of nucleotide (\mavgnt) and nonsynonymous (\mavgaa) mutations in the experimentally evolved P450 populations.}  
For the unselected and monomorphic populations, numbers are the average over all replicates at the indicated generation; for the polymorphic population they are from a random sample, with sampling standard error shown.
\newpage

\centerline{\includegraphics[width=2.5in]{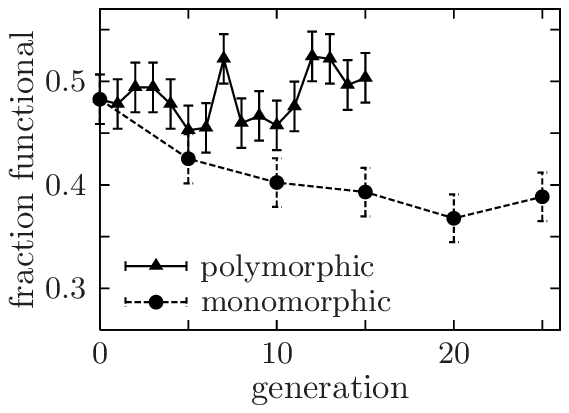}}
\subsection*{\refstepcounter{figurenumber}\label{fig:robustness}Figure \arabic{figurenumber} - The polymorphic population neutrally evolved a higher average mutational robustness than the monomorphic populations.}  
The fraction functional was determined by assaying 435 mutants (average of 1.5 nucleotide mutations per gene).  Error bars show binomial standard error.  For the monomorphic population, numbers are the average over all replicates.
\newpage

\centerline{\includegraphics[width=6.5in]{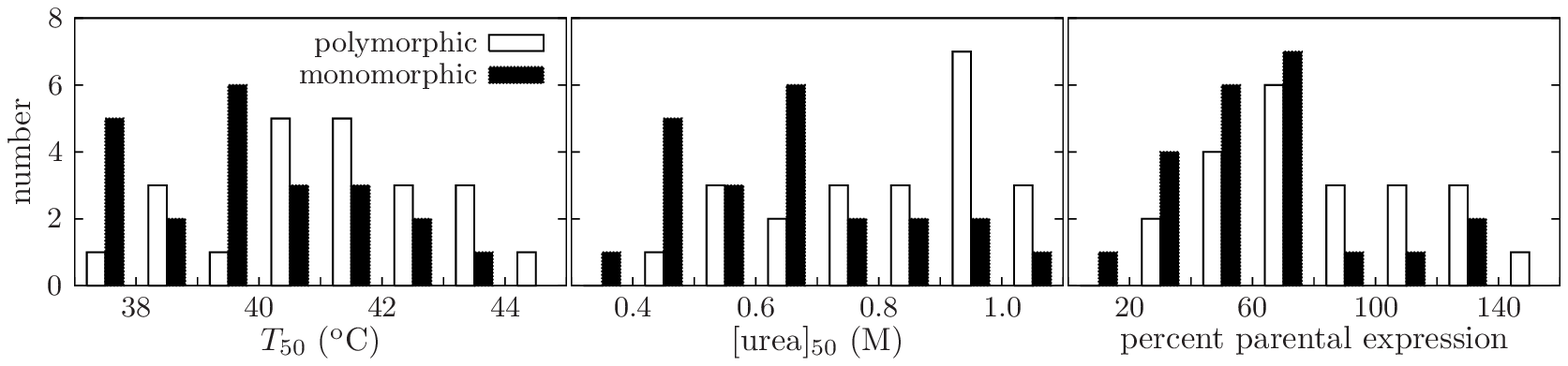}}
\subsection*{\refstepcounter{figurenumber}\label{fig:stabilities_expression}Figure \arabic{figurenumber} - The more mutationally robust proteins are more stable.} 
The P450s from the polymorphic population neutrally evolved higher stability and expression levels than their counterparts from the monomorphic populations.  The histograms show the distributions for the final protein from all monomorphic replicates and for the same number of randomly chosen proteins from the final polymorphic population.  The plots show (left to right) the temperature at which half the protein irreversibly denatured after 10 minutes (\thalf), the urea concentration at which half the protein denatured after 4 hours (\uhalf), and the expression level relative to that of the original parental P450.  The means are significantly different, with unequal variance t-test $P$-values of $0.02$, $0.005$, and $0.04$, respectively.

\newpage

\section*{Tables}
\newcounter{tablenumber}

\subsection*{\refstepcounter{tablenumber}\label{tab:mutspectrum}Table \arabic{tablenumber} - Error-prone PCR nucleotide mutation spectrum.}
Spectrum of nucleotide mutations introduced by the error-prone PCR procedure used in the neutral evolution experiments.  The spectrum was determined by sequencing the four final (generation 12) sequences from the unselected population, since in these sequences the mutations accumulate without constraint.  As has been previously noted for error-prone PCR with \textit{Taq} polymerase~\cite{Bloom2005,Bloom2006,Shafikhani1997}, the nucleotide error spectrum is biased towards certain types of mutations. \par \mbox{}
\par
\mbox{
\begin{tabular}{|l|c|}  \hline
Total nucleotide mutations & 67 \\ \hline
\% synonymous mutations & 25 \\ \hline
Mutation types (\%) & \\
\qquad A \ra T, T \ra A & 19.4 \\
\qquad A \ra C, T \ra G & 1.5 \\
\qquad A \ra G, T \ra C & 64.2 \\
\qquad G \ra A, C \ra T & 4.5 \\
\qquad G \ra C, C \ra G & 0.0 \\
\qquad G \ra T, C \ra A & 1.5 \\
\qquad frameshift & 9.0 \\ \hline
\end{tabular}
}
\newpage

\subsection*{\refstepcounter{tablenumber}\label{tab:evolutionsummary}Table \arabic{tablenumber} - Neutral evolution robustness and mutation data.}
Each row is for a different generation, $T$.  Entries of NA indicate that no measurement was made.  The $\langle m_{nt}\rangle$ and $\langle m_{aa}\rangle$ are the average number of nucleotide mutations and nonsynonymous mutations, respectively.  Numbers in parentheses are total counts over the total samples.  Subscripts indicate the population type: $U$ for unselected, $P$ for polymorphic, and $M$ for monomorphic.  For the unselected and monomorphic populations, numbers represent averages of all replicates.  For the polymorphic population, numbers are for a random sample of functional mutants. \avgFpoly\ and \avgFmono\ are the fraction of functional mutants out of 435 assayed. \par \mbox{}
\par
\mbox{\tiny
\newcommand{\mntp}{$\langle m_{nt} \rangle_P$}
\newcommand{\maap}{$\langle m_{aa} \rangle_P$}
\newcommand{\mntm}{$\langle m_{nt} \rangle_M$}
\newcommand{\maam}{$\langle m_{aa} \rangle_M$}
\newcommand{\mntu}{$\langle m_{nt} \rangle_U$}
\newcommand{\maau}{$\langle m_{aa} \rangle_U$}
\begin{tabular}{|c|c|c|c|c|c|c|c|c|} \hline
$T$ & \mntu & \maau & \mntp & \maap & \mntm & \maam & \avgFpoly & \avgFmono  \\ \hline
0 & 0 & 0 & 0 & 0 & 0 & 0 & 0.48 (210 / 435) & 0.48 (210 / 435) \\ \hline
1 & NA & NA & NA & NA & 0.1 (3 / 22) & 0.3 (6 / 22) & 0.48 (208 / 435) & NA \\ \hline
2 & NA & NA & NA & NA & 0.4 (9 / 22) & 0.8 (17 / 22) & 0.49 (215 / 435) & NA \\ \hline
3 & 5.0 (20 / 4) & 3.5 (14 / 4) & 2.7 (27 / 10) & 1.4 (14 / 10) & 1.0 (23 / 22) & 0.4 (9 / 22) & 0.49 (215 / 435) & NA \\ \hline
4 & NA & NA & NA & NA & 1.5 (32 / 22) &  0.7 (15 / 22) & 0.48 (208 / 435) & NA \\ \hline
5 & NA & NA & NA & NA & 2.2 (48 / 22) & 1.1 (25 / 22) & 0.45 (197 / 435) & 0.43 (185 / 435) \\ \hline
6 & 9.8 (39 / 4) & 7.5 (30 / 4) & 5.5 (55 / 10) & 2.1 (21 / 10) & 2.6 (58 / 22) & 1.4 (31 / 22) & 0.46 (198 / 435) & NA \\ \hline
7 & NA & NA & NA & NA & 3.1 (69 / 22) & 1.8 (39 / 22) & 0.52 (227 / 435) & NA \\ \hline
8 & NA & NA & NA & NA & 3.4 (74 / 22) & 1.8 (40 / 22) & 0.46 (200 / 435) & NA \\ \hline
9 & 13.0 (52 / 4) & 10.3 (41 / 4) & 6.7 (61 / 9) & 3.1 (28 / 9) & 3.7 (82 / 22) & 2.1 (46 / 22) & 0.47 (203 / 435) & NA \\ \hline 
10 & NA & NA & NA & NA & 4.2 (92 / 22) & 2.4 (52 / 22) & 0.46 (199 / 435) & 0.40 (175 / 435) \\ \hline
11 & NA & NA & NA & NA & 4.6 (102 / 22) & 2.5 (56 / 22) & 0.48 (207 / 435) & NA \\ \hline
12 & 16.8 (67 / 4) & 12.5 (50 / 4) & 7.8 (70 / 9) & 3.3 (30 / 9) & 4.9 (107 / 22) & 2.6 (58 / 22) & 0.52 (228 / 435) & NA \\ \hline
13 & NA & NA & NA & NA & 5.0 (110 / 22) & 2.7 (60 / 22) & 0.52 (227 / 435) & NA \\ \hline
14 & NA & NA & NA & NA & 5.3 (116 / 22) & 2.9 (64 / 22) & 0.50 (216 / 435) & NA \\ \hline
15 & NA & NA & 10.3 (227 / 22) & 3.8 (83 / 22) & 5.6 (123 / 22) & 3.0 (67 / 22) & 0.50 (219 / 435) & 0.39 (171 / 435) \\ \hline
16 & NA & NA & NA & NA & 5.8 (127 / 22) & 3.0 (67 / 22) & NA & NA \\ \hline
17 & NA & NA & NA & NA & 6.0 (133 / 22) & 3.1 (69 / 22) & NA & NA \\ \hline
18 & NA & NA & NA & NA & 6.3 (137 / 22) & 3.2 (71 / 22) & NA & NA \\ \hline
19 & NA & NA & NA & NA & 6.3 (138 / 22) & 3.3 (72 / 22) & NA & NA \\ \hline
20 & NA & NA & NA & NA & 6.6 (145 / 22) & 3.4 (75 / 22) & NA & 0.37 (160 / 435) \\ \hline
21 & NA & NA & NA & NA & 6.9 (152 / 22) & 3.6 (79 / 22) & NA & NA \\ \hline
22 & NA & NA & NA & NA & 7.1 (156 / 22) & 3.7 (81 / 22) & NA & NA \\ \hline
23 & NA & NA & NA & NA & 7.2 (158 / 22) & 3.7 (81 / 22) & NA & NA \\ \hline
24 & NA & NA & NA & NA & 7.3 (161 / 22) & 3.8 (83 / 22) & NA & NA \\ \hline
25 & NA & NA & NA & NA &  7.7 (169 / 22) & 4.0 (87 / 22) & NA & 0.39 (169 / 435) \\ \hline
\end{tabular}
}

\newpage

\section*{Additional material}
\newcounter{addfile}

\subsection*{\refstepcounter{addfile}\label{add:parent_P450_R1-11_sequence} Additional file \arabic{addfile} - Sequence of the parent P450 used start neutral evolution.}
FASTA file with sequence of the R1-11 P450 BME used as the neutral evolution parent.  This sequence was isolated after the equilibration evolution.

\subsection*{\refstepcounter{addfile}\label{add:evolution_data}Additional file \arabic{addfile} - Information about sequences from neutral evolution experiments.}
The entries give the name of the mutant, the number of nonsynonymous and nucleotide mutations relative to the R1-11 parent, the \uhalf\ value if measured, the \thalf\ value if measured, the percent of the parental expression level if measured, and then a list of all of the mutations.   Amino acid mutations are numbered in the standard P450 numbering scheme.  The names of the mutants indicate their origin.  Names beginning with ``P-G3'' are randomly chosen functional mutants from generation 3 of the polymorphic population, \textit{etc}.  Names of the form ``P1,'' ``P2,'', \textit{etc.} are the 22 functional mutants that were randomly chosen from the final (generation 15) polymorphic population.  Numbers P5 and P12 are missing because two of the original 24 randomly selected polymorphic population members were randomly chosen to be discarded after it was discovered that two of the 24 monomorphic replicates were contaminated.  Names beginning with ``U1'' indicate that sequences are from the first unselected replicate, \textit{etc.}  Names beginning ``M1'' indicate sequences are from the first monomorphic replicate, \textit{etc}.  Replicates ``M9'' and ``M10'' were discarded due to contamination during the experiment.  For each replicate, we sequenced each new functional mutant.  The last functional mutant after 25 generations represents the final sequence for that replicate, and is given an abbreviated name without the generation suffix.

\subsection*{\refstepcounter{addfile}\label{add:thermostabilities}Additional file \arabic{addfile} - Thermostability measurements.}
Raw data from the \thalf\ thermostability measurements.

\subsection*{\refstepcounter{addfile}\label{add:urea_stabilities}Additional file \arabic{addfile} - Urea stability measurements.}
Raw data from the \uhalf\ thermostability measurements.

\subsection*{\refstepcounter{addfile}\label{add:T50_U50_correlation}Additional file \arabic{addfile} - Correlation of thermal and urea stabilities.}
The \thalf\ and \uhalf\ values are highly correlated.

\subsection*{\refstepcounter{addfile}\label{add:initial_P450_21B3_sequence}Additional file \arabic{addfile} - Sequence of initial P450 used to start equilibration evolution.}
FASTA file with sequence of the 21B3 P450 BM3 heme domain described in \cite{Cirino2003}.  This P450 was used as the initial parent to start the equilibration evolution.

\subsection*{\refstepcounter{addfile}\label{add:equilibration_mutations}Additional file \arabic{addfile} - Mutations accumulated during equilibration evolution.}
The file lists the mutations in the 46 P450 variants selected at the end of the equilibration evolution.  Each line gives the name of the variant, with the prefix indicating whether it came from the R1 or R2 population.  The next entries give the number of nucleotide and nonsynonymous mutations.  All of the individual mutations relative to 21B3 are then listed.  Amino acid mutations are numbered in the standard P450 numbering scheme, with the threonine after the N-terminal methionine given the number one.

\end{bmcformat}
\end{document}